\documentclass[reprint,amsmath,amssymb,aps,pra,floatfix,longbibliography]{revtex4-2}

\usepackage{amsmath,amsfonts,amssymb,graphicx,braket,qcircuit,comment,bm}
\usepackage[]{units} 
\usepackage[T1]{fontenc} 
\usepackage{soul}
\usepackage{hyperref} \hypersetup{colorlinks=true, linkcolor=blue, filecolor=blue, urlcolor=blue, citecolor=blue } 
\usepackage{xcolor}

\DeclareMathOperator{\sgn}{sgn} 

\begin{document} 
%\preprint{Majorana Measurement Simulation}
\title{Measurement-based Simulation of Geometric Gates in Topological Qubits on NISQ Devices} 
\author{Matthew Brooks$^1$}  
\email{matthew.brooks@lps.umd.edu}
\author{Foster Sabatino$^{2,3}$}
\author{Charles Tahan$^{1}$}
\author{Silas Hoffman$^{2,4,5,6}$} 
\affiliation{$^1$Department of Physics, University of Maryland, College Park, MD 20740, USA}
\affiliation{$^2$Department of Physics, University of Florida, Gainesville, FL 32611, USA}
\affiliation{$^3$Department of Physics, University of Central Florida, Orlando, Fl 32816, USA}
\affiliation{$^4$Quantum Theory Project,  University of Florida, Gainesville, FL 32611, USA}
\affiliation{$^5$Laboratory for Physical Sciences, 8050 Greenmead Drive, College Park, Maryland 20740, USA}
\affiliation{$^6$Condensed Matter Theory Center, Department of Physics, University of Maryland, College Park, MD 20742, USA}

%\date{\today}
\begin{abstract}
	
While the adiabatic exchange of Majorana zero modes (MZMs) enables a non-universal set of geometrically protected gates, realising an experimental implementation of MZM braiding remains challenging. In an alternative proposal, charge-parity measurement of two neighboring MZMs supports braiding by teleportation. Moreover, owing to the lack of definitive evidence of MZMs in semiconducting systems, there have been several simulations of MZMs on NISQ devices which more naturally lend themselves to braiding. In this work, measurement-based braiding about MZM Y-junctions are simulated by multi-qubit parity measurements of a logical qubit. Logical single-qubit geometric $S^{(\dagger)}$ and entangling two-qubit braiding operations are shown using two-physical-qubit joint measurements alone, whilst $T^{(\dagger)}$-gates corresponding to a kind of partial-braiding operation, require at least one three-qubit joint measurement. These relatively small scale circuits demonstrate potential applications of both measurement-based geometric gates as well as a measurement-based demonstration of quantum Hamiltonian simulation.

%These relatively small scale circuits offer both novel measurement-based geometric gates as well as a measurement-based demonstration of quantum Hamiltonian simulation.

%Majorana zero modes (MZMs) on a Y-junction may undergo sequences of exchange along each arm of the junction to perform tunable geometric phase gates. Implementation of such gates would offer a universal gate set with topologically protected qubits. However, The adiabatic exchange of MZMs presents a difficult experimental task. Instead, equivalent gates may be enacted by sequences of charge parity measurements between two neighboring MZMz on the Y-junction. This can be simulated on non-topologically protected NISQ devices by sequences of multi-qubit parity measurements on an encoded state. single-qubit geometric $S$ phase gates and entangling two-qubit gates may be shown with two-qubit joint measurements alone, whilst partial phase rotations such as a $T$ gate require at least one three-qubit joint measurement. These relatively small scale circuits offer both novel measurement-based geometric gates as well as a measurement-based demonstration of quantum Hamiltonian simulation.

\end{abstract}
\maketitle

\section{Introduction}

Non-Abelian anyons, such as Majorana zero modes (MZMs) in solid-state systems, could offer a topologically protected platform with which to process and store quantum information\cite{kitaev2001unpaired,kitaev2003fault}. This is due to the complex phases accrued by anyonic particle exchange, allowing for quantum gate implementation by ordered exchange sequences known as braiding. Despite this promise, isolation and braiding of MZMs remains an experimental challenge\cite{yazdani2023hunting}. Additionally, proposed operational schemes for MZM quantum processors tend to lack a universal gate set, omitting the elusive $T$-gate ($\pi/4$ phase gate)\cite{bravyi2005universal}. Although proposals for the implementation of the $T$-gate exist, they lack topological protection\cite{bonderson2010blueprint,clarke2016practical,hoffman2016universal,bravyi2006universal,freedman2006towards}.

An alternative proposal for control of information on a MZM quantum processor is by measurements\cite{xue2013measurement,karzig2017scalable}. This scheme employs joint charge-parity measurements between MZMs coupled by a quantum dot (QD), effectively braiding the MZMs without the need for adiabatic exchange while protecting against quasiparticle-poisoning, the primary source of errors in solid-state MZMs. A notable example of what can be achieved by measurement-based control are geometric phase gates on an encoded four MZM qubit arranged on a Y-junction\cite{karzig2019robust}. This scheme is a measurement-based analogue of the implementation of braided gates\cite{karzig2016universal}, and offers robustness against systematic errors that affect the exchange method. 

Despite the difficulties in realising topologically protected quantum processors, simulations of topological qubits on NISQ devices have been demonstrated. Such simulations include non-Abelian anyonic braiding\cite{stenger2021simulating,xiao2021determining,google2023non,jovanovic2023proposal,sabatino2025simulated}, topological quantum phase transitions\cite{xiao2021determining,chang2022digital,sun2023scalable,shi2023quantum} and chiral edge states\cite{koh2022simulation}. Although NISQ simulation of charge measurement of non-Abelian anyons has been previously discussed\cite{jovanovic2023proposal}, measurement-based control of simulated topological states has not yet been demonstrated. 

In this work, quantum simulation of non-Abelian anyonic geometric phase gates by measurement are discussed and demonstrated on a superconducting NISQ device. By selecting an appropriate topological gauge with an initial entangled state, sequences of two or more parity measurements implement unitary rotations on the encoded state. Additionally, this concept is shown to scale up to perform entangling braiding operations on two encoded MZM qubits coupled by an additional shared MZM Y-junction.  

%These results constitute the first demonstration of geometric measurement-based quantum computing.

\section{Model}

The Hamiltonian of a four-MZM Y-junction is given by

\begin{equation}
    H_{\text{MZM}}=2 i \gamma_0(\bm{\Delta}\cdot\bm{\gamma})
    \label{eq:H_MZM}
\end{equation}

\noindent where $\bm{\gamma}=(\gamma_x,\gamma_y,\gamma_z)$ are the MZMs on each arm of the junction, and $\bm{\Delta}=(\Delta_x,\Delta_y,\Delta_z)$ are the couplings along each arm of the junction\cite{karzig2016universal,karzig2019robust}. Topologically protected geometric gates with such systems were first derived by sequences of exchange given by the tunable coupling vector $\bm{\Delta}(\theta,\phi)=|\bm{\Delta}|(\sin\theta \cos\phi,\sin\theta \sin\phi,\cos\theta)$. The computational subspace of the qubit comprised of these four MZMs is defined by the ground state $a\ket{0}=0$ and excited states $a^{\dagger}\ket{0}=\ket{1}$, with Fermi annihilation (creation) operators are $a^{(\dagger)}=(\gamma_{\hat{\theta}}\pm i\gamma_{\hat{\phi}})/2$ where $\gamma_i=\bm{\gamma}\cdot\hat{e}_i$ and $\hat{e}_i$ is a unit vector defined by the angles associated with an initial coupling vector. By evolving the system couplings so as to trace out an octant on an associated unit sphere of $\bm{\Delta}(\theta,\phi)$, it can be shown that a phase difference between $\ket{0}$ and $\ket{1}$ is gathered given by the solid angle of the octant. This is similar to the Berry phase in a spin system. Therefore, by evolving the system so that $\bm{\Delta}(0,0)\rightarrow \bm{\Delta}(\pi/2,\phi_1)\rightarrow \bm{\Delta}(\pi/2,\phi_2)\rightarrow \bm{\Delta}(0,0)$ traces the solid angle $\Delta\phi=\phi_2-\phi_1$, the phase gate $R_z(\Delta\phi)=e^{-i\Delta\phi \sigma^z/2}$ is performed on the qubit defined by $a^{(\dagger)}=(\gamma_{\hat{x}}\pm i\gamma_{\hat{y}})/2$. This form of the Fermi annihilation (creation) operator is assumed throughout. Here, the total effect of this evolution is to braid $\gamma_x$ and $\gamma_y$ in a tunable way, potentially unlocking $T^{(\dagger)}=R_z(\pm\pi/4)$ operations. Note that $\sgn(\Delta\phi)$ corresponds to which way around the octant given by $|\Delta\phi|$ the system is evolved and $R_z(\Delta\phi)$.

The same octant may be circumnavigated on the unit sphere by measurements\cite{karzig2019robust}. Charge parity projections between $\gamma_0$ and each outer MZMs may be generically given by

\begin{equation}
    \begin{split}
        P_{\text{MZM}}(\theta,\phi)=&1-i\gamma_0 \left(\gamma_x \sin \theta \cos \phi+\gamma_y \sin \theta \sin \phi \right.\\& \left. +\gamma_z \cos \theta\right)
    \end{split}
    \label{eq:P_MZM}
\end{equation}

\noindent so by selecting a sequence of measurement axes $\theta$ and $\phi$ equivalent to that for the exchange sequence, $\gamma_x$ and $\gamma_y$ are braided equivalently, and a measurement-based geometric unitary is applied to the system. In a MZM device, the need to directly probe the charge parity of different pairs of MZMs required coupling between each pair of MZMs via a charge-sensing quantum dot. A viable variant of the MZM hexon architecture to directly probe the necessary charge parities was proposed in Ref.~\cite{karzig2019robust}.

\begin{figure}
    \centering
    \includegraphics[width=0.49\linewidth]{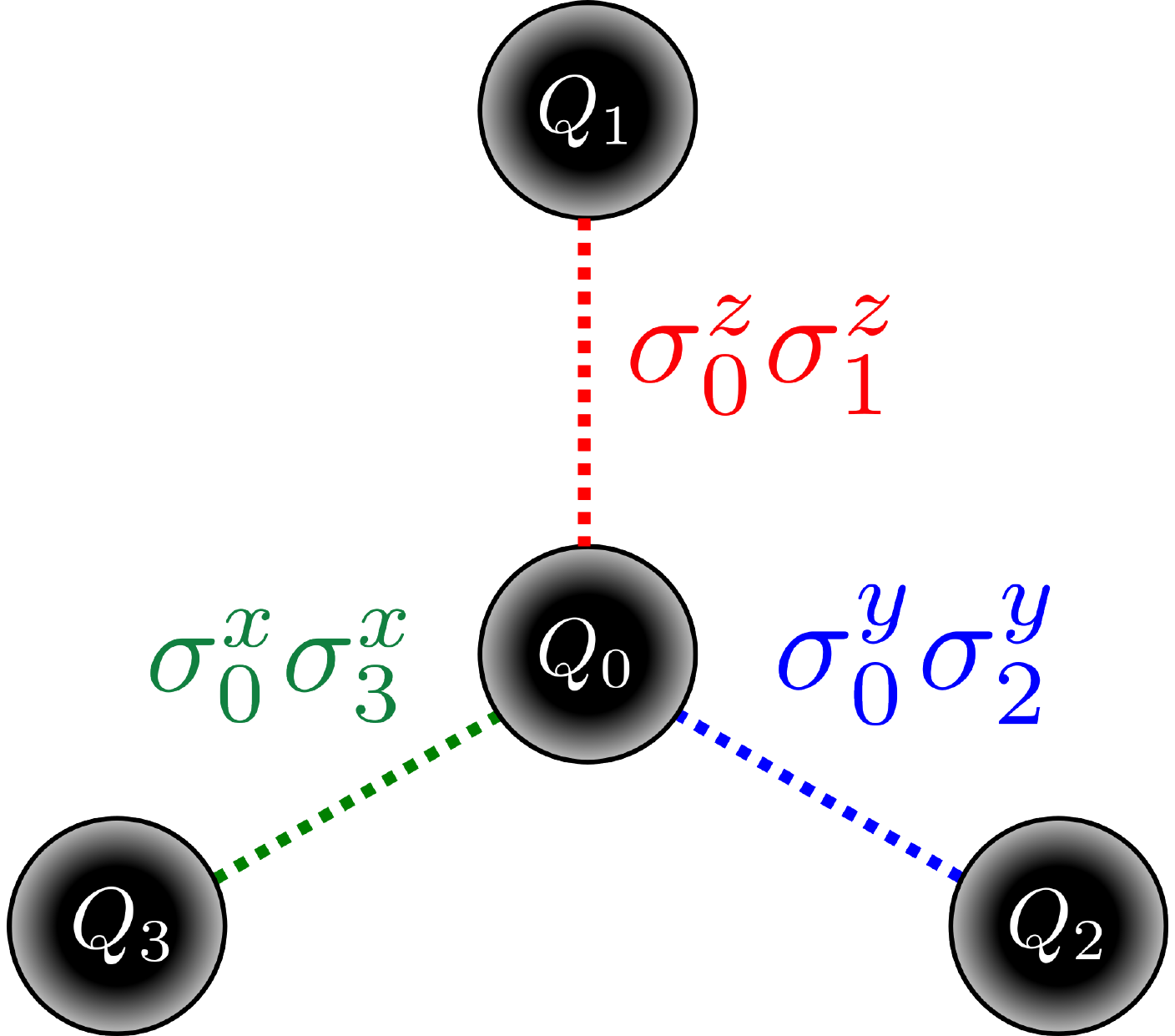}
    \includegraphics[width=0.49\linewidth]{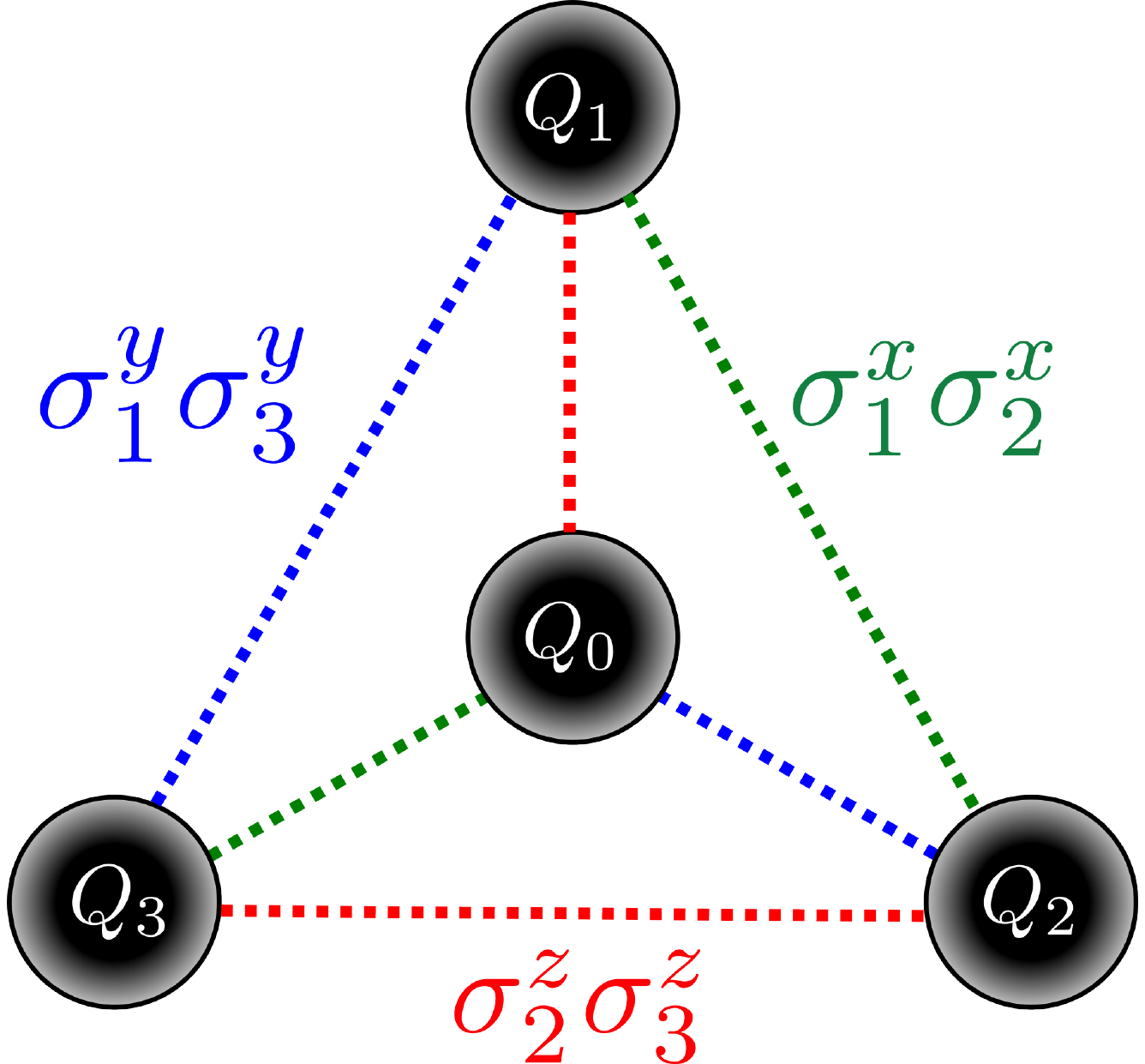}
    \caption{Qubit analogue of a four MZM Y-junction with Kitaev lattice like connectivity. 
    %(b) Qubit analogue of  2 4-MZM Y-junction qubits linked  by a third coupling Y-junction introduced by the addition of two ancilla qubits.
    }
    \label{fig:SingleQubit}
\end{figure}

Suppose instead of 4 MZMs there are 4 qubits, labeled $Q_i$ as in Fig.~\ref{fig:SingleQubit}(a), also arranged as if on a Y-junction. The interactions along each branch of the MZM Y-junction in (\ref{eq:H_MZM}) may be simulated by the following effective Kitaev lattice Hamiltonian

\begin{equation}
    \begin{split}
            H_{\text{sim}}(\theta,\phi) = & \sigma^z_0 \sigma^z_1 \cos \theta + \sigma^y_0 \sigma^y_2 \sin \theta \sin \phi \\ & + \sigma^x_0 \sigma^x_3 \sin \theta \cos \phi
    \end{split}
    \label{eq:FullHam}
\end{equation}

\noindent and projectors simulating those of (\ref{eq:P_MZM}) are given by

\begin{equation}
    P_\pm (\theta,\phi) = \frac{\mathbb{I}\pm H_{\text{sim}}(\theta,\phi)}{2}
    \label{eq:arbObserv}
\end{equation}

\noindent where $\sigma_j^i$ is the $i^{\text{th}}$ Pauli matrix acting on qubit $j$, $\mathbb{I}$ is the identity and $\pm$ dictates measuring the positive or negative eigenstate of the desired observable. Given initialisation to an appropriate logical subspace, a measurement-based geometric phase gate similar to those discussed for a topological system may be demonstrated on a current NISQ processor. A more detailed discussion of this qubitisation method for the Hamiltonian (\ref{eq:H_MZM}) is given in Ref.~\cite{sabatino2025simulated}. 

An appropriate logical subspace is chosen as orthogonal eigenstates to a chosen set of commuting operators 

\begin{equation}
    \begin{split}
        W_1=&\sigma_0^z \sigma_2^x \sigma_3^y\\
        W_2=&\sigma_0^x \sigma_1^y \sigma_2^z
    \end{split}
    \quad\quad
    \begin{split}
        h=&\sigma_0^z \sigma_1^z \\
        %= H_{\text{sim}}(0,0)
        n=&\sigma_2^z \sigma_3^z 
    \end{split}
    \label{eq:Operator_Set}
\end{equation}

\noindent given from the Hamiltonian of a closed four-mode Kitaev lattice, as depicted in Fig.~\ref{fig:SingleQubit}(b). The operators $W_1$ and $W_2$ are two of the three integrals of motion of the closed lattice where $W_3 = n h W_2$. The operator $h$ is chosen to set the gauge of the simulated MZM states, and $n$ differentiates the logical qubit states\cite{sabatino2025simulated}. The chosen logical states are 

\begin{equation}
    \begin{split}
        \tilde{\ket{0}}=&\frac{1}{2}\left(\ket{0101}+\ket{1010}\right)+ \frac{i}{2}\left(\ket{0110}+\ket{1001}\right)\\
        \tilde{\ket{1}}=&\frac{1}{2}\left(\ket{0100}+\ket{1011}\right)- \frac{i}{2}\left(\ket{0111}+\ket{1000}\right)
    \end{split}
    \label{eq:Y1_Logical}
\end{equation}

\noindent both of which are eigenstates with eigenvalue $-1$ of $W_1$, $W_2$ and $h$ whilst $\tilde{\ket{0}}$ is an eigenstate with eigenvalue $-1$ of $n$ and $\tilde{\ket{1}}$ is an eigenstate with eigenvalue $+1$.

The final ingredient for efficient simulation are logical qubit basis measurements for tomography. In the simulated qubit space, the $z$-axis of the Bloch sphere is derived form the Fermi operators as $[a^{\dagger},a]=-i\gamma_x\gamma_y$, therefore defining the $x$-axis as $a+a^{\dagger}=\gamma_x$ and the $y$-axis as $a-a^{\dagger}=\gamma_y$. To translate from the conventional qubit system to the simulated MZM system, we extend the Pauli operators $\sigma_i^\alpha = i \gamma_i^\alpha \gamma_i$, where $u^\alpha_{ij}=i\gamma_i^\alpha \gamma_j^\alpha$ to recover the MZM Hamiltonian (\ref{eq:H_MZM}) by taking $u^z_{01}=u^y_{02}=u^x_{03}=-1$. Equally, by taking $u^z_{23}=-1$, the gauge and qubit basis operators of simulated system (\ref{eq:Operator_Set}), the MZM relations are recovered as $h\rightarrow i\gamma_0\gamma_1\equiv i\gamma_0\gamma_z$ and $n\rightarrow i\gamma_2\gamma_3 \equiv i\gamma_x\gamma_y$ respectively. Therefore, measuring the operator $n$ is equivalent to measuring along the $z$-axis of the simulated Bloch sphere. By extension, with the assumption that the integrals of motion are good quantum numbers $W_1=W_2=-W_3=-1$ and by application of the identity relation $-i\sigma_i^x\sigma_i^y\sigma_i^z=\gamma^x_i\gamma^y_i\gamma^z_i\gamma_i$, it can be shown that the operators $\sigma^y_2\sigma^z_3\rightarrow\gamma_x$ and $\sigma^x_2\rightarrow i\gamma_y$\cite{sabatino2025simulated}. Thus defining the observables to measure along the $x$- and $y$-axes of the simulated MZM qubit respectively.   

\section{single-qubit Rotations}

\subsection{Geometric S-Gate}

The Y-junction configuration of the system of interest naturally lends itself to simulation on superconducting NISQ processors with a heavy-hex lattice architecture, such as the \textit{ibm\_torino} 133 transmon qubit device. Details on this device and demonstration methods used throughout are given in App.~\ref{app:torino}. The simplest measurement-based simulation of a MZM geometric gate braiding operation is the $R_z(\pi/2)$ or $S$-gate. This is achieved by the following measurement sequence: $P_{\pm} (0,0)\rightarrow P_{\pm} (\pi/2,\pi/2)\rightarrow P_{\pm} (\pi/2,0)\rightarrow P_{\pm} (0,0)$, which is equivalent to $\sigma^z_0\sigma^z_1 \rightarrow \sigma^y_0\sigma^y_2 \rightarrow \sigma^x_0\sigma^x_3 \rightarrow \sigma^z_0\sigma^z_1$ two-qubit parity checks. The equivalence of this measurement sequence and an $S$ rotation in the simulated logical qubit space is given in App.~\ref{app:Analytical_Gate}. Note, the first measurement $P_{\pm} (0,0)$ may be omitted as it does not perform any rotation on the initial state. This is because alignment along this axis of the unit sphere is provided by correct initialisation. In previous simulations of geometric MZM operations, an approximate Trotterised form of the adiabatic exchange of MZMs is employed. Here, no such approximations are needed to translate the effect of the charge parity measurements onto the simulated MZM logical qubit.

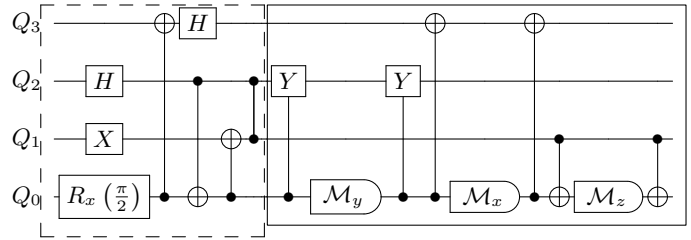
\begin{figure}
    [t]
    \[\Qcircuit @C=0.2em @R=0.6em @!R {
    \lstick{Q_3} & \qw & \targ & \gate{H} & \qw &\qw& \qw   & \qw & \qw & \qw & \qw & \qw & \targ & \qw &  \targ & \qw & \qw & \qw & \qw\\
    \lstick{Q_2} & \gate{H} & \qw & \ctrl{2} & \qw   &\qw & \ctrl{1} &  \qw & \qw & \gate{Y} & \qw &  \gate{Y} & \qw & \qw & \qw & \qw & \qw & \qw & \qw\\
    \lstick{Q_1} &  \gate{X} & \qw & \qw & \targ  &\qw & \ctrl{-1} & \qw & \qw & \qw & \qw & \qw & \qw & \qw & \qw & \ctrl{1} & \qw & \ctrl{1} & \qw\\
    \lstick{Q_0} &  \gate{R_x\left(\frac{\pi}{2}\right)} & \ctrl{-3} & \targ & \ctrl{-1}  &\qw& \qw  & \qw & \qw & \ctrl{-2} & \measureD{\mathcal{M}_y} & \ctrl{-2} & \ctrl{-3} & \measureD{\mathcal{M}_x} & \ctrl{-3} & \targ & \measureD{\mathcal{M}_z} & \targ & \qw
    \gategroup{1}{2}{4}{6}{1.5em}{--} 
    \gategroup{1}{10}{4}{18}{1.5em}{-}
    } \]
    \label{fig:4MZM}
    \caption{Circuit to initialise the required $\tilde{\ket{+}}$ state (dashed box) with 4 qubits and 2-qubit measurement sequence (solid box) to perform the measurement-based geometric $S$-gate in the logical space, and thus simulate a MZM Y-junction operation. $\mathcal{M}_i$ indicates a single qubit measurement on the $i$-axis of the Bloch sphere.} 
\end{figure}

Two-qubit Pauli matrix parity measurements may be implemented either by entangling the data qubits to an ancilla qubit and measuring the ancilla\cite{ustun2023single}, or by performing a measurement circuit decomposing the two-qubit check into a single-qubit measurement directly on the data qubits. The former allows for all the gate measurements to be performed simultaneously at the cost of greater qubit overhead and potentially circuit depth when limited by the connectivity of a given device. Additionally, the performing the measurements simultaneously will not result in the closed octant of the unit sphere simulated, and thus is not applicable here. The latter is a closer translation of the MZM gate simulated and is a more natural fit for the connectivity of the devices used. Fig.~\ref{fig:4MZM} shows the circuit to perform the measurement-based $S$-gate simulation without ancill\ae~ in standard notation. However, the necessity for mid-circuit measurements means substantial idling time for the qubits not being checked at each step of the gate. Idling errors may be addressed by applying dynamical decoupling pulse sequences, in particular an $XY-4$ pulse sequence is used to protect the encoded simulated qubit states. Dephasing errors during readout are not captured in the classical simulation tool of the fake backends provided by \textit{qiskit\_aer}, and so one can expect a reasonable discrepancy between classically simulated results and results from quantum hardware.

Unlike in the initial proposal for the measurement base gate in a MZM Y-junction\cite{karzig2019robust}, here the measurement outcomes of each parity check are probabilistic. Post-selection could be use to focus on the direct implementation of the gates simulated, given only by measuring the $-1$ eigenstate or $1$ output for each parity check. However, like in more conventional measurement-based quantum computing methods\cite{briegel2009measurement,raussendorf2001one,raussendorf2003measurement,brooks2021hybrid,brooks2023quantum}, the different possible outcomes of the gate measurements vary the final state only up to known local corrections. These can either be accounted for in-situ with feed-forward single-qubit rotations applied after the measurement sequence, or by calculating separate density matrices for each gate measurement outcome sequence, and applying the appropriate local corrections after-the-fact. This is known as Pauli-frame tracking. For demonstration and characterisation of a measurement-based gates in this work, the latter correction method is preferred, as it was found to yield more reliable results. Further details on the Pauli-frame corrections needed are given in App.~\ref{app:Pauli_Frame}.

The final ingredient needed to perform the quantum simulations is the choice and preparation of the initial state. While measurements could also be used to initialise the circuits of interest with the knowledge of the gauge choice operator set (\ref{eq:Operator_Set}), a gate based approach to state initialisation is simpler and quicker. All the logical initialisation and logical basis measurement circuits used for the tomography results for the simulation of a 4-MZM system are given in App.~\ref{app:4MZM}.

%To demonstrate the geometric measurement-based $S$-gate with logical basis state tomography, the $\tilde{\ket{+}}=(\tilde{\ket{0}}+\tilde{\ket{1}})/\sqrt{2}$ is initialised with, as shown in Fig.~\ref{fig:4MZM}.

As given in Tab.~\ref{tab:TabFull_Torino}, on the \textit{ibm\_torino} processor, the logical qubit basis, expressed as the process fidelity of the identity operator $I$ is generated with $88.56\%$ fidelity. The measurement-based $S$-gate on the same processor returned output a process fidelity of $71.41\pm0.49\%$, which while overall demonstrating that the measurement-based gate was successful, the difference between the observed fidelity and the classically simulated fidelity of $90.78\pm0.40\%$ is indicative of qubit decay due to long measurement integration times, which is not captured in the simulation package (\textit{qiskit\_aer} using a fake \textit{ibm\_torino} backend). The classical simulations serve as a useful theoretical upper bound to the fidelities observed, and highlight the importance of dynamical decoupling during measurement-based operations with NISQ hardware. The inverse braiding operation, $S^\dagger$ or $R_z(-\pi/2)$, should consist of the same measurement sequence in reverse: $P_{\pm_0} (0,0)\rightarrow P_{\pm_1} (\pi/2,0)\rightarrow P_{\pm_2} (\pi/2,\pi/2)\rightarrow P_{\pm_3} (0,0)$ or $\sigma^z_0\sigma^z_1 \rightarrow \sigma^x_0\sigma^x_3 \rightarrow \sigma^y_0\sigma^y_2 \rightarrow \sigma^z_0\sigma^z_1$, again, omitting the first measurement. On the same device, the $S^\dagger$-gate circuit returned output process fidelity of $74.28\pm0.75\%$, compared to the simulation result of $90.20\pm 0.53\%$. The discrepancy between the simulated and observed fidelities of both the $S^{(\dagger)}$ gates, and the simulated and observed initialisation fidelity indicates a $\sim12\%$ drop in fidelity due to just the measurement sequence, not captured by the \textit{qiskit\_aer} fake backends. These results suggest that a geometric measurement-based gate and simulation of a topological qubit braiding was achieved with good confidence on a NISQ device. All state tomography fidelities from which the process fidelities are derived are reported in App.~\ref{app:State_Fid}.

%Errors accrued in the simulations stem from imperfect gates and measurements, but not idling errors, and so serve as a useful theoretical upper bound to the fidelities measured on \textit{ibm\_torino}

\begin{table}
    [t]
    \centering
    \begin{tabular}{|l||c|c|c|c}
        \hline
         Operation & Av. Depth & Simulation &  \textit{ibm\_torino} \\
         \hline \hline
        $I$ & $25$ & $95.02\%$  & $88.56\%$ \\
        $S$ & $68$ & $90.78\pm 0.40\%$ & $71.41\pm 0.49\%$ \\
        $S^{\dagger}$ & $68$ & $90.20\pm 0.53\%$   &  $74.28\pm0.75\%$ \\
        $T$ & $212$ & $69.30\pm 9.38\%$ & $50.83\pm 8.76\%$ \\
        $T^{\dagger}$ & $220$ & $68.58\pm 10.62\%$  &  $54.91\pm 8.14\%$ \\
        \hline 
        $II$ & $29$ & $91.61\%$ & $81.23\%$ \\
        $R_{xx}\left(\frac{\pi}{2}\right)$ & $76$ & $84.73\pm0.33\%$   &  $45.21\pm0.23\%$ \\
        $R_{xx}\left(-\frac{\pi}{2}\right)$ & $76$ & $84.71\pm0.21\%$ & $47.47\pm0.45\%$  \\
        \hline
    \end{tabular}
    \caption{Table of process fidelities measured in the logical basis for initialisation and measurement-based geometric gate circuits. The average circuit depth of all the process fidelity circuits, \textit{qiskit\_aer} using a fake \textit{ibm\_torino} backend classically simulated fidelities along with the results demonstrated on \textit{ibm\_torino} QPU are given. All demonstrations consist of $2^{15}$ shots, $XY-4$ dynamical decoupling and gate and measurement Pauli-twirling.}
    \label{tab:TabFull_Torino}
\end{table}

\subsection{Geometric T-Gate}

The methods demonstrated may be extended to other geometric phase-gates, specifically the elusive $T^{(\dagger)}$-gate. With the same encoded logical space, the measurement sequence required to rotate the logical state $R_z(\tau)$ is $P_{\pm} (0,0)\rightarrow$ $P_{\pm} (\pi/2,0)\rightarrow$ $P_{\pm} (\pi/2,\tau)\rightarrow$ $P_{\pm} (0,0)$. Here, the challenge is implementing \st{some for of} the following 3-qubit measurement, derived from Eq.~\ref{eq:arbObserv}

\begin{equation}
    P_{\pm} (\pi/2, \tau) = \frac{\mathbb{I}}{2}\pm\frac{1}{2}(\sigma_0^y \sigma_2^y \cos \tau+\sigma_0^x \sigma_3^x \sin \tau)
\end{equation}

\noindent which cannot be generalised as an observable of the tensor product of three Pauli matrices. The form of the two-qubit parity measurement circuits, i.e. a single qubit measurement on one of the relevant data qubits in between relevant entangling rotations, like those show in Fig.~\ref{fig:4MZM} may be extended to three qubits, 

\begin{equation}
    P_{3Q}(\tau)=U_{023}(\tau)\left(\frac{\mathbb{I}\pm R_z(\tau)\sigma_0^x R_z^\dagger(\tau)}{2} \right)U^{\dagger}_{023}(\tau)
    \label{eq:Arb_3Q_M}
\end{equation}

\noindent where $U_{023}(\tau)$ is some entangling three qubit unitary acting on qubits $0$, $2$ and $3$. The entangling unitary can be written as

\begin{equation}
    \begin{split}
            U_{023}(\tau)=\exp\left[\frac{i(\pi + 2\tau)}{4}\left(\mathbb{I}-\sigma_0^z \right)\right.&\\
            \left.\left(\mathbb{I}-\frac{\pi \sigma_1^x-2\tau \sigma_1^x\sigma_2^y}{\pi+2\tau}\right) \right].&
    \end{split}
\end{equation}

In the case $\tau=\pi/4$ the $T$-gate is implemented. The form of the circuit needed to implement $U_{023}(\pi/4)$ is given in Fig.~\ref{fig:T_Rot_Circ}. Compared to the measurement sequence to implement the $S^{(\dagger)}$-gates, the addition of two Toffoli gates and three two-qubit gates each side of the measurement substantially increases the circuit depth. Including the initialisation sequence, the $T$-gate circuit is depth $212$, compared to a depth of $68$ for the $S$-gate, when transpiled for the \textit{ibm\_torino} processor.

This large circuit depth is felt in the results of measurement-based gate as implemented on the \textit{ibm\_torino} processor. The $T$-gate was demonstrated to have a process fidelity of $50.83\pm 8.76\%$ compared to a classically simulated fidelity of $69.30\pm 9.38\%$ while the $T^\dagger$-gate was demonstrated to have a process fidelity of $54.91\pm 8.14\%$ compared to a classically simulated fidelity of $68.58\pm 10.62\%$. These results given in Tab.~\ref{tab:TabFull_Torino}. Note, the reported observed process fidelities for the $T^{(\dagger)}$ gates are to fully characterise the simulations as if they are gates on a simulated qubit, and so are hampered by the averaging of decoherence channels across four separate circuits. The state fidelities observed from each separate circuit demonstration paint a more optimistic picture of the success of the measurement-based gates, with an upper bound fidelity of $71.26\pm6.51\%$ demonstrated from the $T^\dagger \tilde{\ket{i^+}}$ circuit (see App.~\ref{app:State_Fid}). However, the discrepancy between simulated and observed fidelities remains at $\sim12\%$, which is consistent with the $S^{(\dagger)}$ results, further demonstrating that the circuit depth and not the mid circuit measurements are the limiting factor here.

\begin{figure}
    [t]
    \[\Qcircuit @C=0.5em @R=0.5em @!R {
        \lstick{Q_3} & \qw & \targ & \targ & \ctrl{1} & \qw & \qw & \targ & \qw & \qw & \targ & \qw\\
        \lstick{ Q_2} & \qw & \qw & \ctrl{-1} & \ctrl{-1} & \gate{\sqrt{X}^\dagger} & \gate{T^\dagger} & \ctrl{-1} & \gate{T} & \gate{\sqrt{X}} & \ctrl{-1} & \qw \\
        \lstick{Q_0} & \gate{T^\dagger} & \ctrl{-2} & \ctrl{-2} & \qw & \qw & \qw  & \qw  & \qw & \qw & \ctrl{-2} &\qw} \]
    \caption{Quantum circuit implementing $U_{023}(\pi/4)$, as is necessary to implement a simulation of the measurement-based geometric $T^{(\dagger)}$-gates.}
    \label{fig:T_Rot_Circ}
\end{figure}

\section{Geometric Two-Qubit Gates}

\begin{figure}
    [t]
    \centering
    \includegraphics[width=0.89\linewidth]{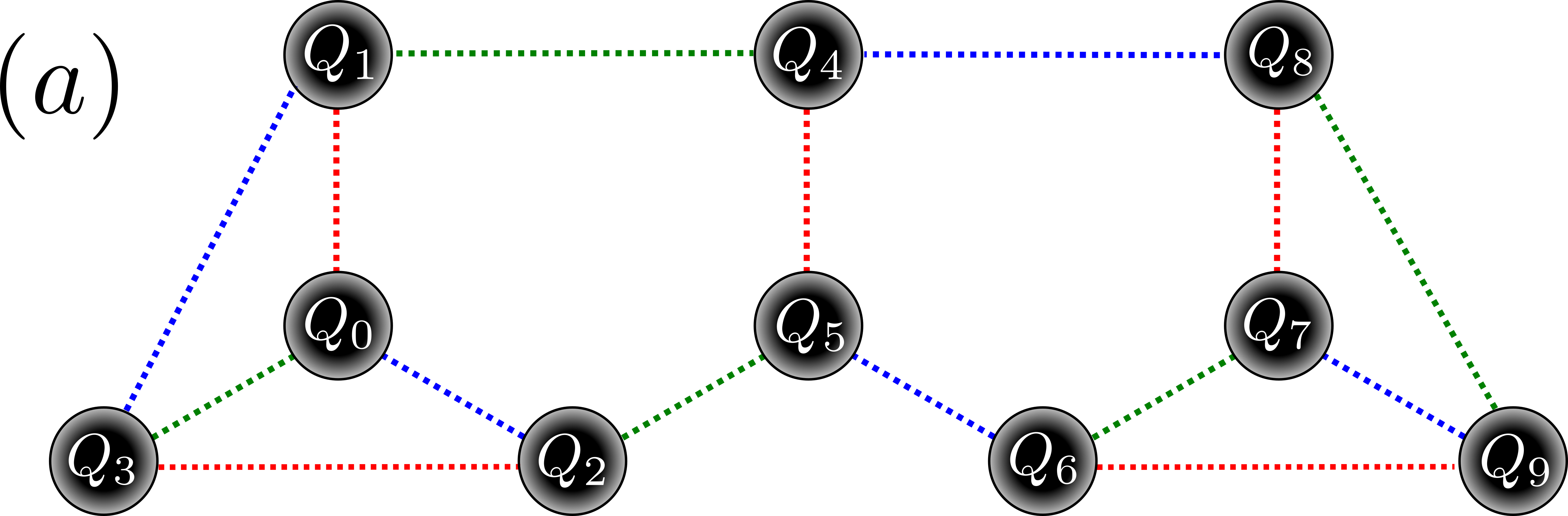}\\
    \vspace{1em}
    \includegraphics[width=0.99\linewidth]{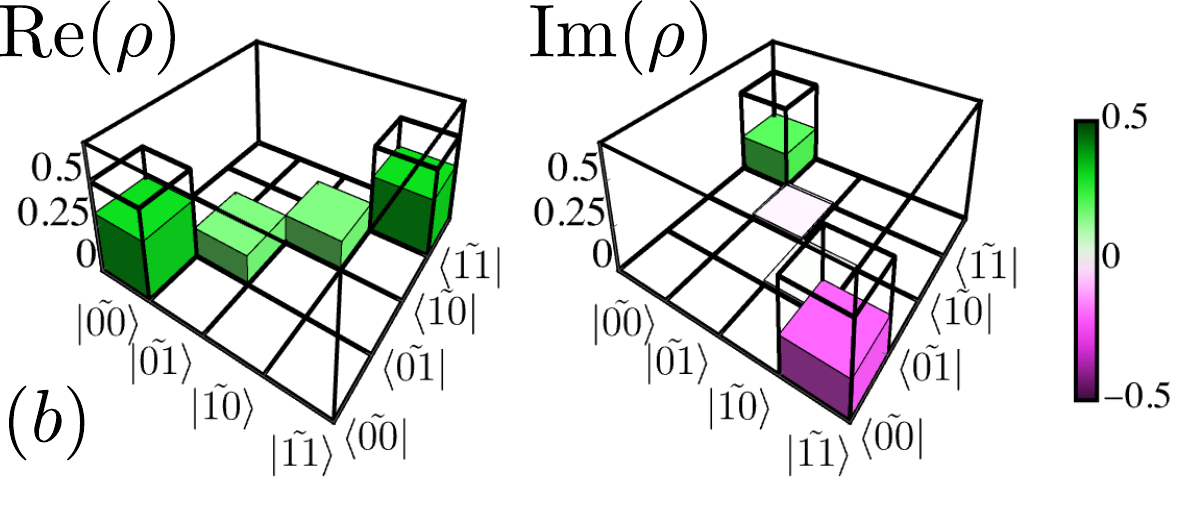}\\
    \caption{(a) Qubit analogue of two simulated Majorana Y-junction qubits labeled $\tilde{Q}_i$ where $i=0,1$, coupled by two ancilla qubits forming a third intermediate Y-junction labeled $\mathcal{A}$. The coloured dashed lines indicate the nature of the coupling between neighbouring qubits in the simulation Hamiltonian, and thus the measurements used to perform operations. Red indicates $\sigma_i^z \sigma_j^z$, blue indicates $\sigma_i^y \sigma_j^y$ and green indicates $\sigma_i^x \sigma_j^x$. (b) Density matrix output in the logical subspace of the simulated $R_{xx}\left(\frac{\pi}{2}\right)$ braiding demonstration performed on the \textit{ibm\_torino} processor, corresponding to a $54.37\%$ output state fidelity.}
    \label{fig:Entangling_gate}
\end{figure}

Thus far, only measurement-based operations with a single Majorana Y-junction qubits have been considered. The methods discussed may be extended to entangling braiding operations of two such qubits. To do so, along with the two Majorana Y-junctions consisting of four modes each, an additional two modes forming a third Y-junction coupling the two-qubits are required. See Fig.~\ref{fig:Entangling_gate} (a). Braiding of MZMs from the two logical qubits is achieved by sequentially measuring the arms of the coupling Y-junction, as in the single-qubit gate operations.

Similarly to the single encoded qubit measurement-based geometric gates simulated with four conventional qubits, entangling operations may be simulated with at least 10 qubits\cite{sabatino2025simulated}. However, due to the extended geometry of the 10 Majorana mode system, the logical states employed in the single encoded qubit demonstrations are not applicable. Additionally, the lower symmetry of the three Y-junctions results in asymmetric simulated logical states between the two encoded qubits labeled $\tilde{Q}_i$ where $i=0,1$. However, the method of deriving the logical states is the same, by finding eigenstates of the following set of commuting gauge selecting operators derived from the Hamiltonian of the 10 mode closed Kitaev lattice

\begin{equation}
    \begin{split}
        W_1&=\sigma_0^y \sigma_1^x \sigma_3^z \\
        W_2&=\sigma_0^z \sigma_2^x \sigma_3^y \\
        W_3&=\sigma_7^x \sigma_8^y \sigma_9^z \\
        W_4&=\sigma_6^y \sigma_7^z \sigma_9^x \\
        W_5&=\sigma_0^x \sigma_1^y \sigma_2^z \sigma_4^y \sigma_5^y
    \end{split}
    \quad\quad
    \begin{split}
        h_{\tilde{Q}_0}&=\sigma_0^z \sigma_1^z \\
        h_{\tilde{Q}_1}&=\sigma_8^z \sigma_7^z \\
        h_{\mathcal{A}}&=\sigma_4^z \sigma_5^z \\
        n_{\tilde{Q}_0}&=\sigma_2^z \sigma_3^z \\
        n_{\tilde{Q}_1}&=\sigma_6^z \sigma_9^z,
    \end{split}
    \label{eq:2qubit_operators}
\end{equation}

\noindent where the $\mp1$ eigenstates of $n_{\tilde{Q}_0}$ and $n_{\tilde{Q}_1}$ define the $\tilde{\ket{0}}(\tilde{\ket{1}})$ of each encoded qubit respectively. The logical basis states for the simulation of the 10-MZM system encoding two topolgical qubits may therefore be given as 

\begin{widetext}
\begin{equation}
    \begin{split}
        \ket{0}^{\tilde Q_0}_L=&\frac{\sqrt{2}}{4}\left(\ket{010101}+\ket{010110}+\ket{101001}+\ket{101010}\right)+\frac{i\sqrt{2}}{4}\left(\ket{011001}+\ket{011010}+\ket{100101}+\ket{100110}\right)\\
        \ket{1}^{\tilde Q_0}_L=&-\frac{\sqrt{2}}{4}\left(\ket{011101}-\ket{011110}+\ket{100001}-\ket{100010}\right)-\frac{i\sqrt{2}}{4}\left(\ket{010001}-\ket{010010}+\ket{101101}-\ket{101110}\right)\\
        \ket{0}^{\tilde Q_1}_L=&\frac{\sqrt{2}(i+1)}{4}\left(\ket{0110}+\ket{1001}\right)- \frac{\sqrt{2}(i-1)}{4}\left(\ket{0101}+\ket{1010}\right)\\
        \ket{1}^{\tilde Q_1}_L=&\frac{\sqrt{2}(i-1)}{4}\left(\ket{0111}+\ket{1000}\right)- \frac{\sqrt{2}(i+1)}{4}\left(\ket{0100}+\ket{1011}\right).\\
    \end{split}
\end{equation}
\label{eq:Y2_logical}
\end{widetext}

\noindent Equally, similarly to the single encoded qubit gates, the operator set (\ref{eq:2qubit_operators}) defines the operator set used for logical basis tomography

\begin{equation}
    \begin{split}
        \sigma_{\tilde{Q}_0}^x &= -i\sigma_2^y \sigma_3^z \sigma_5^z\\
        \sigma_{\tilde{Q}_0}^y &= i\sigma_2^x \sigma_5^z\\
        \sigma_{\tilde{Q}_0}^z &= i n_{\tilde{Q}_0}\\
        %=\sigma_2^z \sigma_3^z\\
    \end{split}
    \quad\quad
    \begin{split}
        \sigma_{\tilde{Q}_1}^x &= i\sigma_6^y\\
        \sigma_{\tilde{Q}_1}^y &= i\sigma_6^x \sigma_9^z\\
        \sigma_{\tilde{Q}_1}^z &=in_{\tilde{Q}_1}.
        %=\sigma_6^z \sigma_9^z.
    \end{split}
\end{equation}

\noindent The notable difference between this set of measurement operators and the operators used to characterise the single encoded qubit operations is the three qubit measurement $\sigma_{\tilde{Q}_0}^x$. All the logical initialisation and logical basis measurement circuits used for the tomography results for the simulation of a 10-MZM system are given in App.~\ref{app:10MZM}. Ideally, when adding more logical qubits to the quantum simulation, measurement of the $W_i$ operators would serve as a more efficient method of initialising. Currently, however, the simulated operations being tested on the available quantum hardware are primarily limited by the mid-circuit measurements required. As such, gate based logical qubit initialisation remains the primary viable method of initialisation, which when scaling beyond two logical qubits, would require larger initial entangled states from initialisation circuits of greater depth. This ultimately would impact the observed simulation fidelities.

%The full form of the logical basis states is given in the appendix, along with the circuit for the measurement-based gate investigated here.

The simulated braiding of the two encoded qubits is achieved by an equivalent measurement sequence to the $S^{(\dagger)}$ gates shown with the single encoded qubit, done on the coupling central Y-junction. Explicitly, by performing a $\sigma^z_4\sigma^z_5 \rightarrow \sigma^y_4\sigma^y_6 \rightarrow \sigma^x_2\sigma^x_4 \rightarrow \sigma^z_4\sigma^z_5$ parity check sequence, the following entangling operation is implemented
    
\begin{equation}
        R_{xx}\left(\frac{\pi}{2}\right)=\exp\left[-\frac{i\pi}{4}\left(\sigma_{\tilde{Q}_0}^x\otimes \sigma_{\tilde{Q}_1}^x\right)\right].
\end{equation}

\noindent Equally, by reversing the measurement sequence, the Hermitian conjugate $R_{xx}\left(-\frac{\pi}{2}\right)$ is implemented instead. In the MZM picture, this operation is equivalent to braiding the MZMs represented by qubits $Q_2$ and $Q_6$ in the qubit analogue show in Fig.~\ref{fig:Entangling_gate} (a).

% The circuits investigated initialise the logical $\ket{0}^{\tilde Q_0}_L \ket{1}^{\tilde Q_0}_L$ state and implements the $R_{xx}\left(\pm \frac{\pi}{2}\right)$ measurement sequences. The resulting states are Bell states in the encoded qubit space
% \begin{equation}
%     R_{xx}\left(\pm \frac{\pi}{2}\right)\ket{0}^{\tilde Q_0}_L \ket{1}^{\tilde Q_0}_L=\frac{1}{\sqrt{2}}\left(\ket{0}^{\tilde Q_0}_L \ket{1}^{\tilde Q_0}_L\mp i\ket{1}^{\tilde Q_0}_L \ket{1}^{\tilde Q_1}_L\right).
% \end{equation}

Similarly to the single-topological-qubit gates, these operations are demonstrated on the \textit{ibm\_torino} processor, with local corrections due to the gate measurement outcomes corrected with classical post-processing. Process fidelities are given in Tab.\ref{tab:TabFull_Torino}. An example circuit of what such a measurement based gate looks like is given in App.~\ref{app:10MZM}. Classical simulations of initialisation and gate fidelities give an expected upper bound in performance of $92.11\%$ and $82.21\pm0.17\%$ respectively. Demonstrations of this on \textit{ibm\_torino} show output state fidelities up to $81.23\%$ for initialisation, $45.21\pm0.23\%$ for the $R_{xx}\left(\frac{\pi}{2}\right)$ gate $47.47\pm0.45\%$ for the $R_{xx}\left(-\frac{\pi}{2}\right)$ gate. Here, it is worth noting that the demonstrated process fidelity results of the entangling operations on our simulated system fall far below our classically simulated fidelities. This is due to a loss of coherence of the larger simulated topological-qubits during the readout integration time of $1.56\mu$s for each mid-circuit measurement, which is not captured by \textit{qiskit\_aer}. Despite the application of an $XY-4$ dynamical decoupling sequence, to further mitigate these losses, more substantial sequences could be employed such as $XY-8$. Lastly, it is also worth noting that, while the process fidelity is a more complete metric of quality of the quantum channel that describes the demonstrated gate, the output state fidelities of each circuit tested to generate such a channel is more optimistic, with highs of up to $68.51\pm0.75\%$ fidelity (see App.~\ref{app:State_Fid}). The density matrix of the result of the $R_{xx}\left(\frac{\pi}{2}\right)\ket{0}^{\tilde Q_0}_L \ket{1}^{\tilde Q_0}_L$ demonstration is shown as in Fig.~\ref{fig:Entangling_gate} (b). Finally, when comparing the discrepancy in the simulated and observed fidelities, the $R_{xx}(\pm\frac{\pi}{2})$ gates suffer a $\sim34\%$ reduction in fidelity due to just the measurement sequence, further demonstrating the susceptibility of the larger encoded logical states to decoherence compared to the $S^{(\dagger)}$ and $T^{(\dagger)}$ gates that suffered a $\sim12\%$ reduction.

%These results successfully demonstrate simulated measurement-based entanglement by braiding of non-Abelian anyonic qubits simulated on a NISQ device.

\section{Conclusion}
In this work, the simulation of measurement-based braiding operations of a non-Abelian anyonic qubit described by a four Majorana Y-junction is demonstrated on a NISQ device. By employing sequential two-qubit parity checks in place of charge parity checks on an appropriately initialised state, simulations of the measurement-based geometric $S^{(\dagger)}$ braiding gates\cite{karzig2019robust} are performed with up to $\sim74\%$ process fidelity. This is extended to show that with the addition of a three qubit non-Pauli measurement, the $T^{(\dagger)}$ gates may be simulated with up to $\sim54\%$ process fidelity. Finally, entangling operations by the braiding of MZMs from two such anyonic qubits was simulated by sequential measurements around an ancilla Y-junction coupling the two-qubits, with output process fidelity up to $\sim47\%$. From our simulations it is clear that there is still a significant error budget associated with mid circuit measurements on current NISQ hardware. As readout in such devices becomes faster and higher fidelity, it would be possible to extend these simulations. For example, ideally one would prefer to initialise the logical qubit space by measuring the associated $W_i$ integral of motion operators as one scales up the number of simulated logical encoded qubits. Similarly, one could use rounds of measurements of the same operators like stabiliser checks, to enforce the logical space and detect errors.

Measurement-based quantum computing is often discussed for its applications to quantum error correction\cite{ustun2023single,briegel2009measurement,raussendorf2001one,raussendorf2003measurement}, or as device specific realisations of operations on a quantum processor\cite{karzig2017scalable,karzig2019robust,brooks2021hybrid,brooks2023quantum}. Here instead, measurement-based quantum simulations have been demonstrated, with little overhead in required qubits, measurements and circuit depths. There is, therefore, scope to scale these simulations to study the effect of measurements on larger anyonic systems such as gapless spin liquids in 2D Kitaev-lattices\cite{kitaev2006anyons,takagi2019concept,broholm2020quantum,hu2024nature}. Finally, the measurement sequences employed in this work are similar to those used in dynamical quantum error correcting codes such as the Floquet code\cite{hastings2021dynamically,davydova2023floquet,paetznick2023performance}, and so the simulation techniques outlined here could be used to benchmark such codes.

\section{Acknowledgments}

We acknowledge helpful discussions with L. Giovia, U. G\"ung\"ord\"u, and Y. Yanay. This work is supported by the U.S. Department of Energy, Office of Science, Basic Energy Sciences under Award No. DE-SC0022089. 

\bibliography{MajMeasBib}

\pagebreak

\appendix

% \section{Physical Qubit State Tomography} \label{app:Tomography}

% \begin{table}
%     [h]
%     \centering
%     \begin{tabular}{l||c|c}
%          & $XY-4$ & $XY-8$ \\
%          \hline \hline
%         $\tilde{\ket{+}}$ & $93.9\%$ & $89.6\%$ \\
%         $S\tilde{\ket{+}}$ & $68.3\pm2.8\%$  &  $73.2\pm1.7\%$ \\
%         $S^{\dagger}\tilde{\ket{+}}$ & $60.6\pm14.3\%$ & $79.8\pm1.4\%$ \\
%     \end{tabular}
%     \caption{Table of average state tomography fidelities and variance of the logical state initialisation and the geometric $S^{(\dagger)}$ measurement-based gates on the \textit{ibm\_lagos} processor, given for circuits employing either $XY-4$ or $XY-8$ dynamical decoupling pulse sequences. All experiments consist of $8000$ shots and applied measurement error mitigation corrections to the tomography measurements.}
%     \label{tab:Lagos_S_tab}
% \end{table}

% \begin{figure}
%     [h]
%     \centering
%     \includegraphics[width=0.99\linewidth]{Figures/HintonPlot_Labeled.png}
%     \caption{Hinton plot of the (a) real and (b) imaginary components of the density matrix of the ideal outcome of $S^\dagger \tilde{\ket{+}}$ compared to the (a) real and (b) imaginary components of the density matrix of the geometric measurement-based gate as implemented on the \textit{ibm\_lagos} processor. White (black) elements are positive (negative) and their magnitude is given by the size of the square. The data shown is of the all $-1$ gate measurement outcomes state from an 8000 shot experiment with $XY-8$ dynamical decoupling.}
%     \label{fig:XY8_Hinton_Gate}
% \end{figure}

\section{\textit{ibm\_torino} Specifications}
\label{app:torino}

\begin{figure}
    [h]
    \centering
    \includegraphics[width=0.99\linewidth]{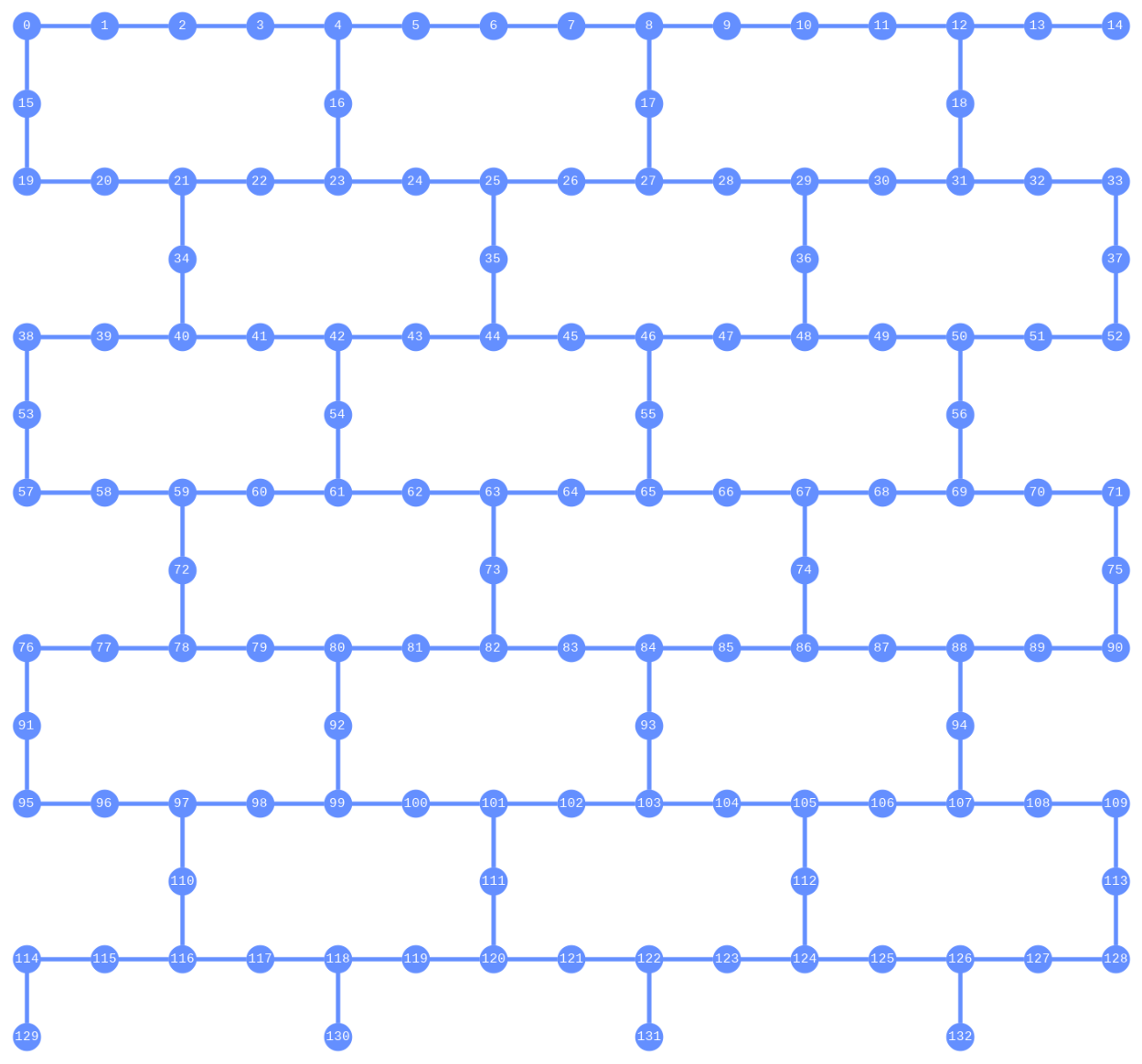}
    \caption{Coupling map of the 133 transmon \textit{ibm\_torino} QPU.}
    \label{fig:Torino_Map}
\end{figure}

All NISQ demonstrations presented in this work were executed on the Heron r1 \textit{ibm\_torino} QPU consisting of 133 transmons arranged on a heavy-hexagonal lattice with tunable couplers connecting interacting qubits, as shown in Fig.~\ref{fig:Torino_Map}. All demonstrations consisted of $2^{15}$ shots sampler primitives, maximum (level 3) transpilation optimisation, $XY-4$ dynamical-decoupling and Pauli-twirling enabled for both gates and measurements. The basis gates for the \textit{ibm\_torino} QPU are $CZ$, $I$, $R_x(\phi)$, $R_z(\phi)$, $R_{xx}(\phi)$, $SX$ and $X$. Readout integration time on the \textit{ibm\_torino} device is $1.56\mu s$. All demonstrations of the simulated 4-MZM, single-logical qubit circuits were executed when the publicly available calibration data was given as: $3.671\times10^{-3}$ median $CZ$ error, $2.862\times10^{-4}$ median $SX$ error, $2.417\times10^{-2}$ median readout error, $185.31\mu s$ median $T_1$ times and $134.13\mu s$ median $T_2$ times. All demonstrations of the simulated 10-MZM, two-logical qubit circuits were executed when the publicly available calibration data was given as: $3.848\times10^{-3}$ median $CZ$ error, $3.053\times10^{-4}$ median $SX$ error, $2.808\times10^{-2}$ median readout error, $185.68\mu s$ median $T_1$ times and $129.66\mu s$ median $T_2$ times.

\section{Measurement-Based Simulated Gates}
\label{app:Analytical_Gate}

Here, the measurement sequences applied in this work to enact the geometric rotations in the logical basis studied are given analytically. This can be shown for the gates $S$ and $T$ from the overlap between the logical states given in (\ref{eq:Y1_Logical}) and the measurement operator (\ref{eq:arbObserv}), for the arbitrary form of the measurement sequence used, $P_{\pm} (0,0)\rightarrow$ $P_{\pm} (\pi/2,0)\rightarrow$ $P_{\pm} (\pi/2,\tau)\rightarrow$ $P_{\pm} (0,0)$, fixing all measurement results to $-1$

\begin{equation}
    \begin{split}
        \bra{\tilde{0}}P_{-} (0,0)P_{-} (\pi/2,\tau)P_{-} (\pi/2,0)P_{-} (0,0)\ket{\tilde{0}}&=e^{-\frac{i \tau}{2}} \\
        \bra{\tilde{0}}P_{-} (0,0)P_{-} (\pi/2,\tau)P_{-} (\pi/2,0)P_{-} (0,0)\ket{\tilde{1}}&=0 \\
        \bra{\tilde{1}}P_{-} (0,0)P_{-} (\pi/2,\tau)P_{-} (\pi/2,0)P_{-} (0,0)\ket{\tilde{0}}&=0 \\
        \bra{\tilde{1}}P_{-} (0,0)P_{-} (\pi/2,\tau)P_{-} (\pi/2,0)P_{-} (0,0)\ket{\tilde{1}}&=e^{\frac{i \tau}{2}}.
    \end{split}
\end{equation}

\noindent Therefore, if $\tau=\pi/2$, an $S$ gate is achieved, and if $\tau=\pi/4$, a $T$ gate is achieved. For all other measurement outcomes, one can show that the desired gate is achieved up to a Pauli correction. Equally, if the measurement sequence is reversed, the following overlaps are derived

\begin{equation}
    \begin{split}
        \bra{\tilde{0}}P_{-} (0,0)P_{-} (\pi/2,0)P_{-} (\pi/2,\tau)P_{-} (0,0)\ket{\tilde{0}}&=e^{\frac{i \tau}{2}} \\
        \bra{\tilde{0}}P_{-} (0,0)P_{-} (\pi/2,0)P_{-} (\pi/2,\tau)P_{-} (0,0)\ket{\tilde{1}}&=0 \\
        \bra{\tilde{1}}P_{-} (0,0)P_{-} (\pi/2,0)P_{-} (\pi/2,\tau)P_{-} (0,0)\ket{\tilde{0}}&=0 \\
        \bra{\tilde{1}}P_{-} (0,0)P_{-} (\pi/2,0)P_{-} (\pi/2,\tau)P_{-} (0,0)\ket{\tilde{1}}&=e^{-\frac{i \tau}{2}}.
    \end{split}
\end{equation}

\noindent which is equivalent to an $S^\dagger$ gate when $\tau=\pi/2$ and $T^\dagger$ gate when $\tau=\pi/4$. Equivalently, from the logcial states given in (\ref{eq:Y2_logical}) for the 10-MZM, two simulated logical qubit space, and extending the projection operator given in (\ref{eq:arbObserv}), the $R_{xx}(\pm\frac{\pi}{2})$ gates may be derived.

\section{Pauli Frame Tracking}
\label{app:Pauli_Frame}

Here, details on the logical corrections needed to account for the probabilistic measurement outcomes of all the measurement-based gate sequences demonstrated in the work are given. In Tab.~\ref{tab:pauli_corr_ST} the corrections applied to the density matrices constructed from state tomography in the 4-MZM, single simulated topological qubit gate demonstrations are given. Here, note that the corrections depend on the direction navigated by the measurement sequences, and not the total phase accumulation of those gate. I.e. the $S$ and $T$ gates have the same local corrections based on gate measurement outcome and equivalently for the $S^\dagger$ and $T\dagger$. The logical corrections to the entangling gates in the 10-MZM, two simulated topological qubit demonstrations are given in Tab.~\ref{tab:pauli_corr_RXX}. In a longer sequence of gates with measurement-dependant Pauli errors, the inclusion of non-Clifford gates like $T^{(\dagger)}$ would add non-Pauli errors to the final result of the circuit. In that case, dynamical circuits that make use of low-latency classical logic, to feed the necessary corrections to the implemented gates into the circuit as it is being executed. While such circuits are possible on current cloud-based IBM quantum hardware, they are currently incompatible with dynamical decoupling, a vital ingredient for mitigating dephasing of the physical qubits during the relatively long mid-circuit measurements, and so were found to not be viable or necessary for the presented simulations.

\begin{table}
    [t]
    \centering
    \begin{tabular}{|l|c|c|}
        \hline
         Measurement & $S/T$ & $S^\dagger/T^\dagger$  \\
         \hline \hline
        000 & $X$  & $Y$   \\
        001 & $Y$  & $X$   \\
        010 & $Y$  & $X$   \\
        011 & $X$  & $Y$   \\
        100 & $I$  & $I$   \\
        101 & $Z$  & $Z$   \\
        110 & $Z$  & $Z$   \\
        111 & $I$  & $I$   \\
        \hline 
    \end{tabular}
    \caption{Table of the logical corrections applied to the constructed density matrices for each 4-MZM, single simulated topological qubit gate demonstrated given by the probabilistic outcome of the gate measurement sequence.}
    \label{tab:pauli_corr_ST}
\end{table}

\begin{table}
    [h]
    \centering
    \begin{tabular}{|l|c|c|}
        \hline
         Measurement & $R_{xx}(\pi/2)$  & $R_{xx}(-\pi/2)$ \\
         \hline \hline
        000 & $IY$  & $XZ$   \\
        001 & $II$  & $II$   \\
        010 & $XZ$  & $IY$   \\
        011 & $XX$  & $XX$   \\
        100 & $XZ$  & $IY$   \\
        101 & $XX$  & $XX$   \\
        110 & $IY$  & $XZ$   \\
        111 & $II$  & $II$   \\
        \hline 
    \end{tabular}
    \caption{Table of the logical corrections applied to the constructed density matrices for each 10-MZM, two simulated topological qubit gate demonstrated given by the probabilistic outcome of the gate measurement sequence.}
    \label{tab:pauli_corr_RXX}
\end{table}

\section{4-MZM Circuits}
\label{app:4MZM}

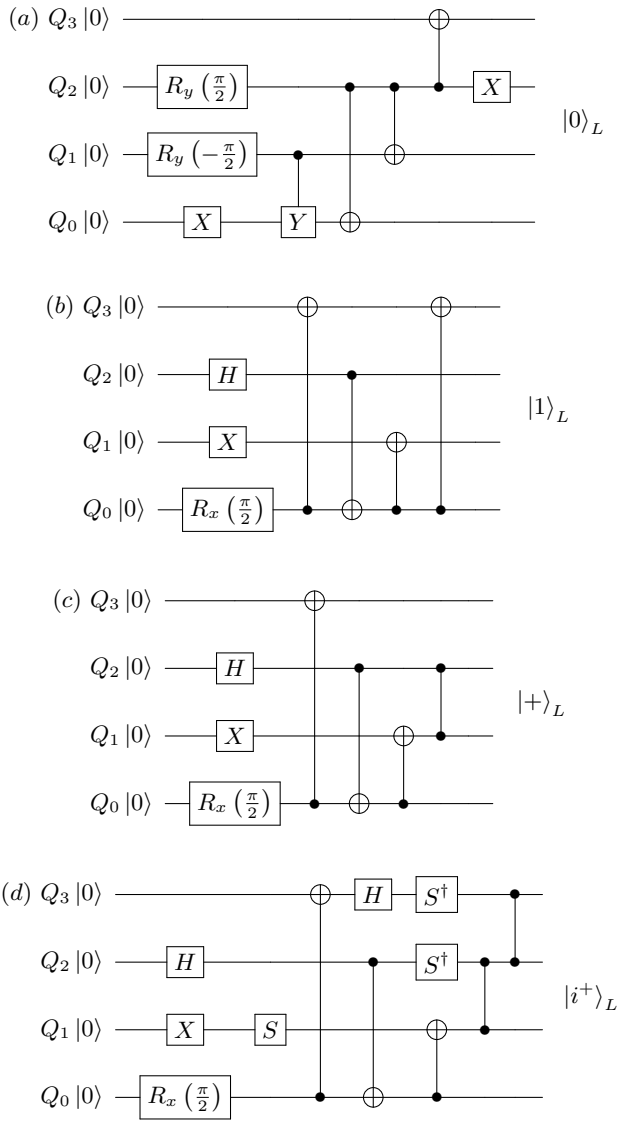
\begin{figure}
    [t]
    \[\Qcircuit @C=1em @R=1em @!R {
        \lstick{(a)\text{        }Q_3\ket{0}}&  \qw & \qw & \qw & \qw & \targ & \qw & \qw \\
        \lstick{Q_2\ket{0}}& \gate{R_y\left(\frac{\pi}{2}\right)} & \qw & \ctrl{2} & \ctrl{1} & \ctrl{-1}& \gate{X} & \qw   \\
        \lstick{Q_1\ket{0}}& \gate{R_y\left(-\frac{\pi}{2}\right)} & \ctrl{1} & \qw & \targ & \qw & \qw & \qw && \raisebox{3em}{$\ket{0}_L$} \\
        \lstick{Q_0\ket{0}}& \gate{X}& \gate{Y}& \targ & \qw & \qw & \qw & \qw  \\
    } \]

    \[\Qcircuit @C=1em @R=1em @!R {
    \lstick{(b)\text{        }Q_3\ket{0}}&  \qw & \targ & \qw & \qw & \targ & \qw & \qw \\
    \lstick{Q_2\ket{0}}& \gate{H} & \qw & \ctrl{2} & \qw & \qw& \qw & \qw   \\
    \lstick{Q_1\ket{0}}& \gate{X} & \qw & \qw & \targ & \qw & \qw & \qw && \raisebox{3em}{$\ket{1}_L$} \\
    \lstick{Q_0\ket{0}}& \gate{R_x\left(\frac{\pi}{2}\right)}& \ctrl{-3}& \targ & \ctrl{-1} & \ctrl{-3} & \qw & \qw  \\
    } \]

    \[\Qcircuit @C=1em @R=1em @!R {
     \lstick{(c)\text{        }Q_3\ket{0}}&  \qw & \targ & \qw & \qw & \qw & \qw & \qw \\
     \lstick{Q_2\ket{0}}& \gate{H} & \qw & \ctrl{2} & \qw & \control \qw & \qw & \qw   \\
     \lstick{Q_1\ket{0}}& \gate{X} & \qw & \qw & \targ & \ctrl{-1} & \qw & \qw && \raisebox{3em}{$\ket{+}_L$} \\
    \lstick{Q_0\ket{0}}& \gate{R_x\left(\frac{\pi}{2}\right)}& \ctrl{-3}& \targ & \ctrl{-1} & \qw & \qw & \qw  \\
    } \]

    \[\Qcircuit @C=1em @R=1em @!R {
    \lstick{(d)\text{        }Q_3\ket{0}}&  \qw& \qw  & \targ & \gate{H} & \gate{S^\dagger} & \qw & \control \qw  & \qw \\
    \lstick{Q_2\ket{0}}& \gate{H}& \qw  & \qw & \ctrl{2} & \gate{S^\dagger} & \control \qw & \ctrl{-1}& \qw   \\
    \lstick{Q_1\ket{0}}& \gate{X}& \gate{S}  & \qw & \qw & \targ & \ctrl{-1} & \qw & \qw && \raisebox{3em}{$\ket{i^+}_L$} \\
    \lstick{Q_0\ket{0}}& \gate{R_x\left(\frac{\pi}{2}\right)}& \qw &  \ctrl{-3}& \targ & \ctrl{-1} & \qw & \qw & \qw  \\  
    } \]
\caption{Circuits to initialize the required logical states (a) $\ket{0}_L$, (b) $\ket{1}_L$, (c) $\ket{+}_L$ and (d) $\ket{i^+}_L$, for process tomography for the 4-MZM, simulated topological qubit demonstrations.} 
\label{fig:Logical_Init_Y1}
\end{figure}

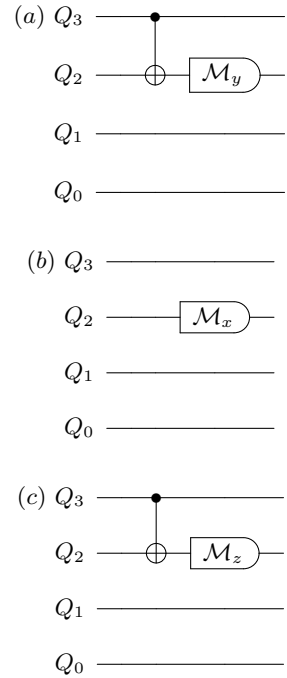
\begin{figure}
    [t]
    \[\Qcircuit @C=1em @R=1em @!R {
    \lstick{(a)\text{        }Q_3}& \qw & \ctrl{1} & \qw & \qw  \\
    \lstick{Q_2}& \qw & \targ & \measureD{\mathcal{M}_y} & \qw  \\
    \lstick{Q_1}& \qw & \qw & \qw & \qw  \\
    \lstick{Q_0}& \qw & \qw & \qw & \qw  \\
    } \]

    \[\Qcircuit @C=1em @R=1em @!R {
    \lstick{(b)\text{        }Q_3}& \qw & \qw & \qw & \qw  \\
    \lstick{Q_2}& \qw & \qw & \measureD{\mathcal{M}_x} & \qw  \\
    \lstick{Q_1}& \qw & \qw & \qw & \qw  \\
    \lstick{Q_0}& \qw & \qw & \qw & \qw  \\
    } \]

    \[\Qcircuit @C=1em @R=1em @!R {
    \lstick{(c)\text{        }Q_3}& \qw & \ctrl{1} & \qw & \qw  \\
    \lstick{Q_2}& \qw & \targ & \measureD{\mathcal{M}_z} & \qw  \\
    \lstick{Q_1}& \qw & \qw & \qw & \qw  \\
    \lstick{Q_0}& \qw & \qw & \qw & \qw  \\
    } \]
\caption{Circuits to measure in the (a) $x$ ($\sigma_2^y\sigma_3^z$), (b) $y$ ($\sigma_2^x$) and (c) $z$ ($\sigma_2^z\sigma_3^z$) logical basis for process tomography for the 4-MZM, simulated topological qubit demonstrations.} 
\label{fig:Logical_Measure_Y1}
\end{figure}

All the necessary circuits needed to initilise (Fig.~\ref{fig:Logical_Init_Y1}) and read out the logical basis (Fig.~\ref{fig:Logical_Measure_Y1}) of the simulated 4-MZM system are given here. To generate the appropriate choi-matricies used to calculate the process fidelities of each of the measurement-based gates demonstrated, each state-tomography on each measurement-based gate is performed efficiently by measuring only the logical qubit basis, for each initial state given in Fig.~\ref{fig:Logical_Init_Y1}.

\section{State Fidelity Results}
\label{app:State_Fid}

Here, the state tomography results for each demonstration from which the process fidelities reported in Tab.~\ref{tab:TabFull_Torino}. To construct the Choi-matricies used to calculate the reported process fidelities, density matrices from demonstrations with initial states $\ket{0}$, $\ket{1}$, $\ket{+}=(\ket{0}+\ket{1})/\sqrt{2}$ and $\ket{+i}=(\ket{0}+i\ket{1})/\sqrt{2}$ for each logical qubit are needed. Tab.~\ref{tab:ST_state} details all simulated and demonstrated state fidelities for the all simulated 4-MZM, single-topological-qubit measurement-based gates investigated.

\begin{table}
    [h]
    \centering
    \begin{tabular}{|l|c|c|}
        \hline
         Circuit & Simulation  & \textit{ibm\_torino} \\
         \hline \hline
        $S\tilde{\ket{0}}$ & $94.07\pm0.32\%$  & $75.97\pm0.51\%$  \\
        $S\tilde{\ket{1}}$ & $95.78\pm0.34\%$  & $79.04\pm0.72\%$  \\
        $S\tilde{\ket{+}}$ & $80.75\pm0.49\%$  & $71.22\pm0.47\%$  \\
        $S\tilde{\ket{i^+}}$ & $80.37\pm0.53\%$  & $75.04\pm0.69\%$  \\
        $S^\dagger\tilde{\ket{0}}$ & $95.20\pm0.16\%$  & $80.20\pm0.82\%$  \\
        $S^\dagger\tilde{\ket{1}}$ & $96.33\pm0.12\%$  & $79.13\pm0.58\%$  \\
        $S^\dagger\tilde{\ket{+}}$ & $82.27\pm0.53\%$  & $77.67\pm0.74\%$  \\
        $S^\dagger\tilde{\ket{i^+}}$ & $78.18\pm0.30\%$  & $71.78\pm0.50\%$  \\
        \hline
        $T\tilde{\ket{0}}$ & $82.07\pm3.96\%$  & $69.16\pm3.73\%$  \\
        $T\tilde{\ket{1}}$ & $80.15\pm4.96\%$  & $64.76\pm4.18\%$  \\
        $T\tilde{\ket{+}}$ & $76.08\pm8.75\%$  & $65.23\pm6.42\%$  \\
        $T\tilde{\ket{i^+}}$ & $81.30\pm5.77\%$  & $69.69\pm6.54\%$  \\
        $T^\dagger\tilde{\ket{0}}$ & $82.21\pm3.61\%$  & $69.22\pm3.61\%$  \\
        $T^\dagger\tilde{\ket{1}}$ & $82.14\pm3.78\%$  & $70.31\pm2.81\%$  \\
        $T^\dagger\tilde{\ket{+}}$ & $79.41\pm7.98\%$  & $69.14\pm6.09\%$  \\
        $T^\dagger\tilde{\ket{i^+}}$ & $79.11\pm8.26\%$  & $71.26\pm6.51\%$  \\
        \hline 
    \end{tabular}
    \caption{Table of state fidelities measured in the logical basis for all 4-MZM, single-topological-qubit measurement-based gates investigated. The \textit{qiskit\_aer} using a fake \textit{ibm\_torino} backend classically simulated fidelities along with the results demonstrated on \textit{ibm\_torino} QPU are given. All demonstrations consist of $2^{15}$ shots, $XY-4$ dynamical decoupling and gate and measurement Pauli-twirling.}
    \label{tab:ST_state}
\end{table}

\begin{table}
    [h]
    \centering
    \begin{tabular}{|l|c|c|}
        \hline
         Circuit & Simulation  & \textit{ibm\_torino} \\
         \hline \hline
        $R_{xx}(\pi/2)\ket{0}^{\tilde Q_0}_L \ket{0}^{\tilde Q_1}_L$ & $88.51\pm0.28\%$  & $54.37\pm0.47\%$  \\
        $R_{xx}(\pi/2)\ket{0}^{\tilde Q_0}_L \ket{1}^{\tilde Q_1}_L$ & $89.79\pm0.33\%$  & $63.07\pm0.33\%$  \\
        $R_{xx}(\pi/2)\ket{0}^{\tilde Q_0}_L \ket{+}^{\tilde Q_1}_L$ & $85.90\pm0.33\%$  & $56.04\pm0.43\%$  \\
        $R_{xx}(\pi/2)\ket{0}^{\tilde Q_0}_L \ket{i^+}^{\tilde Q_1}_L$ & $87.62\pm0.28\%$  & $55.71\pm0.84\%$  \\
            $R_{xx}(\pi/2)\ket{1}^{\tilde Q_0}_L \ket{0}^{\tilde Q_1}_L$ & $90.00\pm0.51\%$  & $64.16\pm0.70\%$  \\
        $R_{xx}(\pi/2)\ket{1}^{\tilde Q_0}_L \ket{1}^{\tilde Q_1}_L$ & $91.41\pm0.42\%$  & $67.47\pm0.69\%$  \\
        $R_{xx}(\pi/2)\ket{1}^{\tilde Q_0}_L \ket{+}^{\tilde Q_1}_L$ & $86.77\pm0.20\%$  & $58.76\pm0.51\%$  \\
        $R_{xx}(\pi/2)\ket{1}^{\tilde Q_0}_L \ket{i^+}^{\tilde Q_1}_L$ & $89.63\pm0.40\%$  & $64.65\pm0.55\%$  \\
                $R_{xx}(\pi/2)\ket{+}^{\tilde Q_0}_L \ket{0}^{\tilde Q_1}_L$ & $88.27\pm0.25\%$  & $50.47\pm0.34\%$  \\
        $R_{xx}(\pi/2)\ket{+}^{\tilde Q_0}_L \ket{1}^{\tilde Q_1}_L$ & $87.84\pm0.43\%$  & $61.01\pm0.62\%$  \\
        $R_{xx}(\pi/2)\ket{+}^{\tilde Q_0}_L \ket{+}^{\tilde Q_1}_L$ & $84.66\pm0.72\%$  & $51.61\pm0.59\%$  \\
        $R_{xx}(\pi/2)\ket{+}^{\tilde Q_0}_L \ket{i^+}^{\tilde Q_1}_L$ & $88.43\pm0.31\%$  & $51.99\pm0.65\%$  \\
                $R_{xx}(\pi/2)\ket{i^+}^{\tilde Q_0}_L \ket{0}^{\tilde Q_1}_L$ & $85.43\pm0.36\%$  & $52.87\pm0.82\%$  \\
        $R_{xx}(\pi/2)\ket{i^+}^{\tilde Q_0}_L \ket{1}^{\tilde Q_1}_L$ & $89.72\pm0.23\%$  & $61.13\pm0.90\%$  \\
        $R_{xx}(\pi/2)\ket{i^+}^{\tilde Q_0}_L \ket{+}^{\tilde Q_1}_L$ & $87.72\pm0.30\%$  & $51.70\pm0.62\%$  \\
        $R_{xx}(\pi/2)\ket{i^+}^{\tilde Q_0}_L \ket{i^+}^{\tilde Q_1}_L$ & $85.39\pm0.30\%$  & $52.99\pm0.56\%$  \\
        \hline
    \end{tabular}
    \caption{Table of state fidelities measured in the logical basis for the 10-MZM, two-topological-qubit measurement-based $R_{xx}(\pi/2)$ gate. The \textit{qiskit\_aer} using a fake \textit{ibm\_torino} backend classically simulated fidelities along with the results demonstrated on \textit{ibm\_torino} QPU are given. All demonstrations consist of $2^{15}$ shots, $XY-4$ dynamical decoupling and gate and measurement Pauli-twirling.}
    \label{tab:RXX_state}
\end{table}

\begin{table}
    [h]
    \centering
    \begin{tabular}{|l|c|c|}
        \hline
         Circuit & Simulation  & \textit{ibm\_torino} \\
         \hline \hline
        $R_{xx}(-\pi/2)\ket{0}^{\tilde Q_0}_L \ket{0}^{\tilde Q_1}_L$ & $86.21\pm0.36\%$  & $54.01\pm0.44\%$  \\
        $R_{xx}(-\pi/2)\ket{0}^{\tilde Q_0}_L \ket{1}^{\tilde Q_1}_L$ & $91.27\pm0.40\%$  & $62.73\pm0.49\%$  \\
        $R_{xx}(-\pi/2)\ket{0}^{\tilde Q_0}_L \ket{+}^{\tilde Q_1}_L$ & $88.75\pm0.34\%$  & $60.79\pm0.61\%$  \\
        $R_{xx}(-\pi/2)\ket{0}^{\tilde Q_0}_L \ket{i^+}^{\tilde Q_1}_L$ & $85.58\pm0.17\%$  & $57.88\pm0.42\%$  \\
        $R_{xx}(-\pi/2)\ket{1}^{\tilde Q_0}_L \ket{0}^{\tilde Q_1}_L$ & $89.89\pm0.22\%$  & $63.04\pm0.67\%$  \\
        $R_{xx}(-\pi/2)\ket{1}^{\tilde Q_0}_L \ket{1}^{\tilde Q_1}_L$ & $92.53\pm0.41\%$  & $64.67\pm1.00\%$  \\
        $R_{xx}(-\pi/2)\ket{1}^{\tilde Q_0}_L \ket{+}^{\tilde Q_1}_L$ & $89.51\pm0.23\%$  & $68.51\pm0.75\%$  \\
        $R_{xx}(-\pi/2)\ket{1}^{\tilde Q_0}_L \ket{i^+}^{\tilde Q_1}_L$ & $90.88\pm0.24\%$  & $63.97\pm0.49\%$  \\
        $R_{xx}(-\pi/2)\ket{+}^{\tilde Q_0}_L \ket{0}^{\tilde Q_1}_L$ & $84.69\pm0.18\%$  & $57.51\pm0.78\%$  \\
        $R_{xx}(-\pi/2)\ket{+}^{\tilde Q_0}_L \ket{1}^{\tilde Q_1}_L$ & $86.50\pm0.38\%$  & $60.96\pm0.29\%$  \\
        $R_{xx}(-\pi/2)\ket{+}^{\tilde Q_0}_L \ket{+}^{\tilde Q_1}_L$ & $85.55\pm0.31\%$  & $61.06\pm0.58\%$  \\
        $R_{xx}(-\pi/2)\ket{+}^{\tilde Q_0}_L \ket{i^+}^{\tilde Q_1}_L$ & $85.22\pm0.47\%$  & $57.31\pm0.51\%$  \\
        $R_{xx}(-\pi/2)\ket{i^+}^{\tilde Q_0}_L \ket{0}^{\tilde Q_1}_L$ & $88.33\pm0.45\%$  & $55.75\pm0.68\%$  \\
        $R_{xx}(-\pi/2)\ket{i^+}^{\tilde Q_0}_L \ket{1}^{\tilde Q_1}_L$ & $91.52\pm0.27\%$  & $61.34\pm0.46\%$  \\
        $R_{xx}(-\pi/2)\ket{i^+}^{\tilde Q_0}_L \ket{+}^{\tilde Q_1}_L$ & $89.82\pm0.12\%$  & $56.79\pm0.40\%$  \\
        $R_{xx}(-\pi/2)\ket{i^+}^{\tilde Q_0}_L \ket{i^+}^{\tilde Q_1}_L$ & $86.96\pm0.24\%$  & $51.95\pm0.79\%$  \\
        \hline
    \end{tabular}
    \caption{Table of state fidelities measured in the logical basis for the 10-MZM, two-topological-qubit measurement-based $R_{xx}(-\pi/2)$ gate. The \textit{qiskit\_aer} using a fake \textit{ibm\_torino} backend classically simulated fidelities along with the results demonstrated on \textit{ibm\_torino} QPU are given. All demonstrations consist of $2^{15}$ shots, $XY-4$ dynamical decoupling and gate and measurement Pauli-twirling.}
    \label{tab:RXXM_state}
\end{table}

\pagebreak

\section{10-MZM Circuits}
\label{app:10MZM}

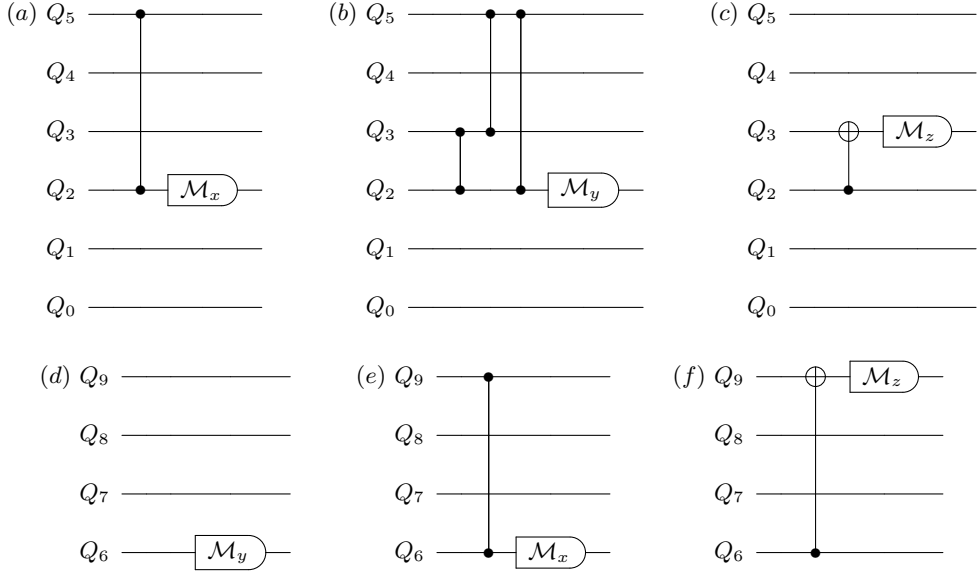
\begin{figure*}
    [b]

    \[\Qcircuit @C=1em @R=1em @!R {
    \lstick{(a)\text{        }Q_5}& \qw &\ctrl{3} &\qw &\qw &&&&&&   \lstick{(b)\text{        }Q_5}& \qw &\qw &\ctrl{2} &\ctrl{3} &\qw&\qw &&&&&&   \lstick{(c)\text{        }Q_5}& \qw &\qw &\qw &\qw   \\
    \lstick{Q_4}& \qw &\qw &\qw  &\qw &&&&&& \lstick{Q_4}& \qw &\qw &\qw  &\qw &\qw&\qw &&&&&&  \lstick{Q_4}& \qw &\qw &\qw  &\qw  \\
    \lstick{Q_3}& \qw &\qw &\qw &\qw &&&&&&  \lstick{Q_3}& \qw &\ctrl{1} &\ctrl{-2} &\qw &\qw&\qw &&&&&&   \lstick{Q_3}& \qw &\targ &\measureD{\mathcal{M}_z} &\qw   \\
     \lstick{Q_2}& \qw &\ctrl{-3} &\measureD{\mathcal{M}_x}  &\qw &&&&&& \lstick{Q_2}& \qw &\ctrl{-1} &\qw  &\ctrl{-3}&\measureD{\mathcal{M}_y}  &\qw &&&&&&  \lstick{Q_2}& \qw &\ctrl{-1} &\qw  &\qw  \\
     \lstick{Q_1}& \qw &\qw &\qw  &\qw &&&&&& \lstick{Q_1}& \qw &\qw &\qw  &\qw &\qw&\qw &&&&&& \lstick{Q_1}& \qw &\qw &\qw  &\qw  \\
     \lstick{Q_0}& \qw &\qw &\qw  &\qw &&&&&& \lstick{Q_0}& \qw &\qw &\qw  &\qw &\qw&\qw &&&&&& \lstick{Q_0}& \qw &\qw &\qw  &\qw  \\
    } \]

    \[\Qcircuit @C=1em @R=1em @!R {
    \lstick{(d)\text{        }Q_9}& \qw &\qw &\qw &\qw &&&&&& \lstick{(e)\text{        }Q_9}& \qw &\ctrl{3} &\qw &\qw &&&&&&  \lstick{(f)\text{        }Q_9}& \qw &\targ &\measureD{\mathcal{M}_z}&\qw   \\
    \lstick{Q_8}& \qw &\qw &\qw  &\qw &&&&&& \lstick{Q_8}& \qw &\qw &\qw  &\qw &&&&&&  \lstick{Q_8}& \qw &\qw &\qw  &\qw  \\
    \lstick{Q_7}& \qw &\qw &\qw &\qw &&&&&& \lstick{Q_7}& \qw &\qw &\qw &\qw &&&&&&  \lstick{Q_7}& \qw &\qw &\qw &\qw   \\
     \lstick{Q_6}& \qw &\qw &\measureD{\mathcal{M}_y}  &\qw&&&&&&  \lstick{Q_6}& \qw &\ctrl{-3} &\measureD{\mathcal{M}_x}  &\qw &&&&&&  \lstick{Q_6}& \qw &\ctrl{-3} &\qw  &\qw  \\
    } \]

\caption{Circuits to measure in the (a) $x_0$ ($\sigma_2^x\sigma_5^z$), (b) $y_0$ ($\sigma_2^y\sigma_3^z\sigma_5^z$), (c) $z_0$ ($\sigma_2^z\sigma_3^z$), (d) $x_1$ ($\sigma_6^y$), (e) $y_1$ ($\sigma_6^x\sigma_9^z$) and (f) $z_1$ ($\sigma_6^z\sigma_9^z$) logical basis for process tomography for the 10-MZM for logical $\tilde{Q}_0$ and $\tilde{Q}_1$, 2 simulated topological qubit demonstrations.} 
\label{fig:Logical_Measure_Y2_Q0}
\end{figure*}

All the necessary circuits needed to initilise (Fig.~\ref{fig:Logical_Init_Y2_Q0}) and read out the logical basis (Fig.~\ref{fig:Logical_Measure_Y2_Q0}) of the simulated 10-MZM system are given here for the simulated logical qubits $\tilde{Q}_0$ and $\tilde{Q}_1$. As in App.~\ref{app:4MZM}, the appropriate combinations of initial logical input states, with tomography measurements performed in the simulated logical basis allows for the efficient probing of the process fidelities of the measurement-based gates demonstrated.

\begin{figure*}
    [t]
    \[\Qcircuit @C=1em @R=1em @!R {
    \lstick{(a)\text{        }Q_5\ket{0}}& \gate{H}& \ctrl{1} & \qw &\qw &\qw & \qw &\qw &\qw &&&&&&&&&&  \lstick{(b)\text{        }Q_5\ket{0}}& \gate{H}&\gate{Z}& \ctrl{1} & \qw &\qw &\qw & \qw   \\
    \lstick{Q_4\ket{0}}& \gate{X}& \targ & \qw &\qw &\qw & \qw &\qw &\qw &&&&&&&&&&  \lstick{Q_4\ket{0}}& \gate{X}& \qw & \targ & \qw &\qw &\qw & \qw   \\
    \lstick{Q_3\ket{0}}& \gate{X}& \gate{S^\dagger}& \gate{Y} &\targ& \qw & \qw&\qw &\qw && \raisebox{-3em}{$\ket{0}^{q_0}_L$} &&&&&&&& \lstick{Q_3\ket{0}}& \qw &  \gate{Y} &\targ& \qw & \qw&\qw &\qw && \raisebox{-3em}{$\ket{1}^{q_0}_L$} \\
     \lstick{Q_2\ket{0}}& \gate{H} & \qw & \ctrl{-1} &\qw & \targ & \gate{Z}&\qw &\qw &&&&&&&&&& \lstick{Q_2\ket{0}}& \gate{H}  & \ctrl{-1} &\qw & \targ & \gate{Z}&\qw &\qw  \\
     \lstick{Q_1\ket{0}}&  \gate{H} & \qw & \qw &\ctrl{-2} & \ctrl{-1} & \ctrl{1} & \gate{Y} &\qw&&&&&&&&&&  \lstick{Q_1\ket{0}}&  \gate{H} & \qw &\ctrl{-2} & \ctrl{-1} & \ctrl{1} & \gate{Y} &\qw  \\
     \lstick{Q_0\ket{0}}&  \qw & \qw &\qw &\qw &\qw & \targ &\qw &\qw &&&&&&&&&& \lstick{Q_0\ket{0}}&  \qw  &\qw &\qw &\qw & \targ &\qw &\qw &
    } \]

    \[\Qcircuit @C=1em @R=1em @!R {
    \lstick{(c)\text{        }Q_5\ket{0}}& \gate{H}&\gate{Z}& \ctrl{1} & \qw &\qw &\ctrl{2} & \ctrl{3} & \qw &&&&&&&&&&   \lstick{(d)\text{        }Q_5\ket{0}}& \gate{H}&\gate{Z}& \ctrl{1}  &\qw &\ctrl{2} & \ctrl{3} & \qw   \\
    \lstick{Q_4\ket{0}}& \gate{X}& \qw & \targ & \qw &\qw &\qw & \qw & \qw &&&&&&&&&&  \lstick{Q_4\ket{0}}& \gate{X}& \qw & \targ  &\qw &\qw & \qw & \qw  \\
    \lstick{Q_3\ket{0}}& \gate{H}& \gate{S} & \gate{Y} &\targ& \gate{Z} & \ctrl{-2}&\qw &\qw && \raisebox{-3em}{$\ket{+}^{q_0}_L$} &&&&&&&& \lstick{Q_3\ket{0}}& \gate{H} & \gate{Y} &\targ& \gate{Z} & \ctrl{-2}&\qw &\qw && \raisebox{-3em}{$\ket{i^+}^{q_0}_L$} \\
     \lstick{Q_2\ket{0}}& \gate{H} & \qw & \ctrl{-1} &\qw & \targ & \qw &\ctrl{-3} &\qw &&&&&&&&&&   \lstick{Q_2\ket{0}}& \gate{H}  & \ctrl{-1} &\qw & \targ & \qw &\ctrl{-3} &\qw  \\
     \lstick{Q_1\ket{0}}&  \gate{H} & \qw & \qw &\ctrl{-2} & \ctrl{-1} & \ctrl{1} & \gate{Y} &\qw &&&&&&&&&&  \lstick{Q_1\ket{0}}&  \gate{H} & \qw &\ctrl{-2} & \ctrl{-1} & \ctrl{1} & \gate{R_y(-\pi)} &\qw  \\
     \lstick{Q_0\ket{0}}&  \qw & \qw &\qw &\qw &\qw & \targ &\qw &\qw &&&&&&&&&& \lstick{Q_0\ket{0}}&  \qw  &\qw &\qw &\qw & \targ &\qw &\qw &
    } \]
    
    \[\Qcircuit @C=1em @R=1em @!R {
     \lstick{(e)\text{        }Q_9\ket{0}}&  \gate{X} & \qw & \qw & \targ & \qw & \qw & \qw &&&&&&&&&&     \lstick{(f)\text{        }Q_9\ket{0}}&  \qw & \qw & \qw & \targ & \qw & \qw & \qw & \qw \\
     \lstick{Q_8\ket{0}}& \gate{X} & \targ & \qw & \qw & \targ& \qw & \qw &&&&&&&&&&   \lstick{Q_8\ket{0}}& \gate{X} & \targ & \qw & \qw &\qw & \targ& \qw & \qw   \\
     \lstick{Q_7\ket{0}}& \gate{H} & \qw & \ctrl{1} & \qw & \qw & \qw & \qw && \raisebox{3em}{$\ket{0}^{q_1}_L$} &&&&&&&& \lstick{Q_7\ket{0}}& \gate{H} & \qw & \ctrl{1} & \qw & \gate{Z} & \qw & \qw & \qw && \raisebox{3em}{$\ket{1}^{q_1}_L$} \\
    \lstick{Q_6\ket{0}}& \gate{\sqrt{X}}& \ctrl{-2}& \targ & \ctrl{-3} & \ctrl{-2} & \qw & \qw &&&&&&&&&&  \lstick{Q_6\ket{0}}& \gate{\sqrt{X}}& \ctrl{-2}& \targ & \ctrl{-3} & \qw & \ctrl{-2} & \qw & \qw  \\
  } \]

    \[\Qcircuit @C=1em @R=1em @!R {
    \lstick{(g)\text{        }Q_9\ket{0}}&  \gate{H} & \qw & \qw &\qw & \targ & \qw & \qw & \ctrl{2} & \qw & \qw &&&&&&&&&& \lstick{(h)\text{        }Q_9\ket{0}}&  \gate{H} & \gate{S^\dagger} & \qw &\qw & \targ & \qw & \qw & \ctrl{2} & \qw & \qw \\
    \lstick{Q_8\ket{0}}& \gate{X} & \targ & \qw &\qw & \qw & \targ& \qw & \qw & \gate{Z} & \qw &&&&&&&&&&   \lstick{Q_8\ket{0}}& \gate{X} & \targ & \qw &\qw & \qw & \targ& \qw & \qw & \gate{Z} & \qw   \\
    \lstick{Q_7\ket{0}}& \gate{H} & \qw & \ctrl{1} &\gate{S} & \qw & \qw & \qw & \ctrl{-2} & \qw & \qw  && \raisebox{3em}{$\ket{+}^{q_1}_L$} &&&&&&&& \lstick{Q_7\ket{0}}& \gate{H} & \qw & \ctrl{1} &\gate{S} & \qw & \qw & \qw & \ctrl{-2} & \qw & \qw  && \raisebox{3em}{$\ket{i^+}^{q_1}_L$} \\
    \lstick{Q_6\ket{0}}& \gate{\sqrt{X}}& \ctrl{-2}& \gate{Y} &\qw& \ctrl{-3} & \ctrl{-2} & \qw & \qw & \qw & \qw &&&&&&&&&&  \lstick{Q_6\ket{0}}& \gate{\sqrt{X}}& \ctrl{-2}& \gate{Y} &\qw& \ctrl{-3} & \ctrl{-2} & \qw & \qw & \qw & \qw  \\
    } \]
\caption{Circuits to initialize the required logical states (a) $\ket{0}^{\tilde{Q}_0}_L$, (b) $\ket{1}^{\tilde{Q}_0}_L$, (c) $\ket{+}^{\tilde{Q}_0}_L$, (d) $\ket{i^+}^{\tilde{Q}_0}_L$, (e) $\ket{0}^{\tilde{Q}_1}_L$, (f) $\ket{1}^{\tilde{Q}_1}_L$, (g) $\ket{+}^{\tilde{Q}_1}_L$ and (h) $\ket{i^+}^{\tilde{Q}_1}_L$, for process tomography for the 10-MZM, 2 simulated topological qubit demonstrations.} 
\label{fig:Logical_Init_Y2_Q0}
\end{figure*}

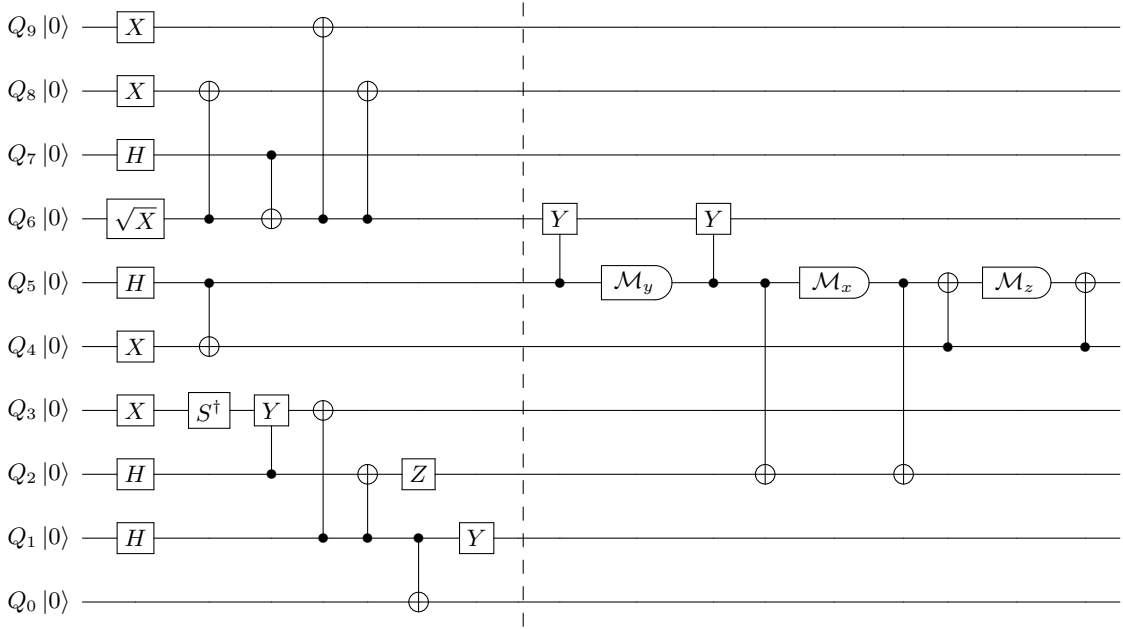
\begin{figure*}
    [h]
      \[\Qcircuit @C=1em @R=1em @!R {
        \lstick{Q_9 \ket{0}}& \gate{X}  & \qw & \qw & \targ&\qw & \qw & \qw  & \barrier[-1.5em]{9} \qw & \qw & \qw & \qw & \qw & \qw & \qw & \qw & \qw & \qw & \qw &\\
       \lstick{Q_8 \ket{0}}& \gate{X}  & \targ & \qw & \qw & \targ&\qw &\qw & \qw & \qw & \qw & \qw & \qw & \qw & \qw & \qw & \qw & \qw & \qw &\\
        \lstick{Q_7 \ket{0}}& \gate{H}  & \qw & \ctrl{1} & \qw&\qw & \qw & \qw & \qw & \qw & \qw & \qw & \qw & \qw & \qw & \qw & \qw & \qw & \qw &\\
        \lstick{Q_6 \ket{0}}& \gate{\sqrt{X}} & \ctrl{-2} & \targ & \ctrl{-3} & \ctrl{-2}&\qw & \qw & \qw & \gate{Y} & \qw & \gate{Y} & \qw & \qw & \qw & \qw & \qw & \qw & \qw &\\
        \lstick{Q_5 \ket{0}}& \gate{H}  & \ctrl{1}& \qw & \qw & \qw & \qw & \qw & \qw & \ctrl{-1} & \measureD{\mathcal{M}_y} & \ctrl{-1} & \ctrl{3} & \measureD{\mathcal{M}_x} & \ctrl{3} & \targ & \measureD{\mathcal{M}_z} & \targ & \qw &\\
        \lstick{Q_4 \ket{0}}& \gate{X}  & \targ & \qw & \qw & \qw & \qw & \qw & \qw & \qw & \qw & \qw & \qw & \qw & \qw & \ctrl{-1} & \qw & \ctrl{-1} & \qw &\\
        \lstick{Q_3 \ket{0}}& \gate{X}&\gate{S^\dagger} & \gate{Y} & \targ & \qw & \qw &\qw & \qw & \qw & \qw & \qw & \qw & \qw & \qw & \qw & \qw & \qw & \qw &\\
        \lstick{Q_2 \ket{0}}& \gate{H}& \qw  & \ctrl{-1} & \qw & \targ & \gate{Z} & \qw & \qw & \qw & \qw & \qw & \targ & \qw & \targ & \qw & \qw & \qw & \qw &\\
        \lstick{Q_1 \ket{0}}& \gate{H}& \qw  & \qw & \ctrl{-2} & \ctrl{-1} & \ctrl{1} & \gate{Y} & \qw & \qw & \qw & \qw & \qw & \qw & \qw & \qw & \qw & \qw & \qw &\\
        \lstick{Q_0 \ket{0}}& \qw & \qw & \qw  & \qw & \qw &  \targ & \qw & \qw & \qw & \qw & \qw & \qw & \qw & \qw & \qw & \qw & \qw & \qw &
  } \]
  \label{fig:10_Qubit_Circuit_Measure}
  \caption{10 qubit circuit to perform the $R_{xx}\left(\frac{\pi}{2}\right)\ket{0}^{\tilde Q_0}_L \ket{1}^{\tilde Q_0}_L$ simulated braiding gate. To the left of the dashed barrier is the simulation basis initialisation circuit, to the right is the sequence of measurements responsible for the simulated rotation.}
  %The qubit labels $Q_i$ correspond to the qubit layout given in Fig.~\ref{fig:Entangling_gate}
\end{figure*}

The full circuit to perform the $R_{xx}\left(\frac{\pi}{2}\right)$ simulated braiding gates on the logical input state $\ket{0}^{\tilde Q_0}_L \ket{1}^{\tilde Q_0}_L$ is given in Fig.~\ref{fig:10_Qubit_Circuit_Measure}. Note that to perform a $R_{xx}\left(-\frac{\pi}{2}\right)\ket{0}^{\tilde Q_0}_L \ket{1}^{\tilde Q_0}_L$ simulated braiding gates, one need only re-arrange the order of measurements.

\end{document}